\pdfoutput=1
\documentclass[11pt,british]{article}
\usepackage[T1]{fontenc}
\usepackage[latin9]{inputenc}
\usepackage[a4paper]{geometry}
\usepackage{babel}

\usepackage{amstext}
\usepackage{graphicx}
\usepackage{amssymb}
\usepackage{esint}
\usepackage[unicode=true, 
 bookmarks=true,bookmarksnumbered=false,bookmarksopen=false,
 breaklinks=false,pdfborder={0 0 1},backref=false,colorlinks=false]
 {hyperref}
\hypersetup{pdftitle={Filtering out the cosmological constant in the Palatini formalism of modified gravity},
 pdfauthor={Florian Bauer},
 pdfsubject={A filter for large cosmological constants},
 pdfkeywords={Cosmological constant, vacuum energy, modified gravity, Palatini formalism}}
\begin{document}

\title{Filtering out the cosmological constant in the Palatini formalism
of modified gravity}

\author{\textbf{Florian Bauer}\vspace*{0.5cm}
 \\
{\small High Energy Physics Group, Dept.\ ECM, and Institut de
Ci\`encies del Cosmos~(ICC)}\\
{\small Universitat de Barcelona, Mart\'i i Franqu\`es 1, E-08028
Barcelona, Catalonia, Spain}\\
{\small Email: }\texttt{\small \href{mailto:fbauerphysik@eml.cc}{fbauerphysik@eml.cc}}}

\date{{ }}
\maketitle
\begin{abstract}
According to theoretical physics the cosmological constant~(CC) is
expected to be much larger in magnitude than other energy densities
in the universe, which is in stark contrast to the observed Big Bang
evolution. We address this old CC problem not by introducing an extremely
fine-tuned counterterm, but in the context of modified gravity in
the Palatini formalism. In our model the large CC term is filtered
out, and it does not prevent a standard cosmological evolution. We
discuss the filter effect in the epochs of radiation and matter domination
as well as in the asymptotic de~Sitter future. The final expansion
rate can be much lower than inferred from the large CC without using
a fine-tuned counterterm. Finally, we show that the CC filter works
also in the Kottler (Schwarzschild-de~Sitter) metric describing a
black hole environment with a CC compatible to the future de~Sitter
cosmos.
\end{abstract}

\section{Introduction}

The starting point of this work is a CC or equivalently a vacuum energy
density~$\Lambda$ of enormous magnitude. This expectation is suggested
by contributions to the CC coming from phase transitions in the early
universe, zero-point energy in quantum field theory or even from quantum
gravity. In general, all these parts are of different magnitude and
probably unrelated to each other. Hence, the sum~$\Lambda$ of all
terms is dominated by the largest contribution. Since other energy
sources dilute with the expansion of the universe, the CC will eventually
take control over the cosmos. Depending on its sign the CC would induce
in the very early universe either a Big Crunch or an eternal de~Sitter
phase with a very high Hubble rate~$H\propto\Lambda$. Obviously,
the standard Big Bang evolution does not happen in this case.

The simplest way to avoid this problem is the introduction of a CC
counterterm~$\Lambda_{\text{ct}}$, which makes the sum~$|\Lambda+\Lambda_{\text{ct}}|$
smaller than the currently observed critical energy density~$\rho_{c0}\sim10^{-47}\,\text{GeV}^{4}$.
For concreteness let us assume that~$\Lambda\sim-M_{\text{ew}}^{4}$
were related to the electroweak phase transition at the energy scale~$M_{\text{ew}}\sim10^{2}\,\text{GeV}$.
Then the counterterm must be extraordinarily close to~$(-\Lambda)$
requiring an enormous amount of fine-tuning,\begin{equation}
\left|1+\frac{\Lambda_{ct}}{\Lambda}\right|<\left|\frac{\rho_{c0}}{\Lambda}\right|\sim10^{-55}.\label{eq:counterterm}\end{equation}
Apart from the fine-tuning of the classical counterterm, the situation
is even more involved when quantum corrections are included, cf.~Ref.~\cite{Bauer:2010wj}
for an elaborated discussion in the context of the electroweak sector
of the standard model of particles. Moreover, the problem worsens
when~$\Lambda$ is dominated by higher energy scales, possibly originating
from grand unified theories where~$\Lambda\sim(10^{16}\,\text{GeV})^{4}$
or quantum gravity with~$\Lambda\sim(10^{19}\,\text{GeV})^{4}$ for
instance. Summing up, the fine-tuning of the CC is considered to be
one of the most severe problems in theoretical physics~\cite{Weinberg:1988cp,Bertolami:2009nr}.
In addition, the current accelerated expansion of the universe~\cite{Spergel:2006hy,Knop:2003iy,Riess:2004nr}
can be explained very well by a tiny CC of the same magnitude as the
energy density of matter, giving rise to the so-called coincidence
problem. For the latter problem many explanations have been proposed~\cite{Carroll:2000fy,Peebles:2002gy,Padmanabhan:2002ji,Nojiri:2006ri,Copeland:2006wr},
which induce late-time accelerated expansion. However, most of these
models tacitly assume that the large~$\Lambda$ has been fine-tuned
away and thus they do not address the big CC problem.

Without fine-tuning we have to accept the existence of the presumed
huge CC, and we have to find a way to neutralise its effects in order
to obtain a reasonable cosmological evolution. Along this line, several
proposals have been made, e.g.\ relaxation models for a large CC
in the context of matter with an inhomogeneous equation of state~(EOS)~\cite{Stefancic:2008zz},
or in the LXCDM framework~\cite{Grande:2006nn} with a variable cosmological
term \cite{Bauer:2009ke,Bauer:2009jk}, see also~\cite{Bonanno:2001hi,Nobbenhuis:2004wn,Barr:2006mp,Diakonos:2007au,Klinkhamer:2007pe,Batra:2008cc}.
Removing or filtering out vacuum energy has been investigated e.g.\
in Refs.~\cite{Dvali:2007kt,Patil:2008sp,Demir:2009be,Hassan:2010ys},
and it is a feature in unimodular gravity~\cite{Unruh:1988in,Henneaux:1989zc,Ng:1990rw,Ng:1990xz,Smolin:2009ti}.
Recently, a CC relaxation model has been discussed in the context
of modified gravity with an action functional~$f(R,G)$ involving
the Ricci scalar~$R$ and the Gau\ss-Bonnet invariant~$G$ in the
metric formalism~\cite{Bauer:2009ea,Bauer:2010wj}, where the action
is varied with respect to the metric~$g_{ab}$ only.

In this work, we also consider a modified gravity model with an action
functional~$f$ in terms of the Ricci scalar~$R$ and the squared
Ricci tensor~$Q=R_{ab}R^{ab}$. However, here we apply the Palatini
formalism, where the metric~$g_{ab}$ and the connection~$\Gamma_{bc}^{a}$
are treated independently by the variation principle. In contrast,
the metric formalism requires from the beginning that \begin{equation}
\Gamma_{bc}^{a}[g]=\frac{1}{2}g^{ad}(g_{dc,b}+g_{bd,c}-g_{bc,d})\end{equation}
is the Levi-Civita connection of~$g_{ab}$, whereas the Palatini
connection depends also on the matter sector. As we will show in this
paper, this allows the construction of a filter for a large CC, and
the results will be similar in effect to unimodular gravity. However,
in our setup the CC is not eliminated completely, but it appears in
suppressed corrections. Furthermore, we investigate the filter effect
from the early universe till the asymptotic future in addition to
black hole environments. It turns out that finite vacuum energy shifts
originating e.g.\ from phase transitions can be neutralised, too.

There is an interesting conceptual difference to the CC relaxation
models in the metric formalism, where the large vacuum energy is not
filtered out from the total energy content, but gravity is modified
such that~$\Lambda$ does not induce large curvatures. This happens
on the level of the Einstein equations, i.e.\ by solving differential
equations to obtain concrete low-curvature solutions. In the Palatini
framework of this paper, we will show that the large CC can be removed
already in an algebraic way before solving differential equations.
Moreover, the latter are only of second order, whereas the metric
version of modified gravity generally involves a higher differential
order, signalling the existence of new degrees of freedom, which can
be the source of new instabilities and other problems. More differences
between both formalisms will become visible in the forthcoming discussions
in this paper.

Some work on the Palatini formalism can be found e.g.\ in Refs.~\cite{Vollick:2003ic,Allemandi:2004wn,Sotiriou:2005xe,Barausse:2007ys,Li:2007xw,Sotiriou:2008rp},
and comparisons with the metric and other formalisms were made in
Refs.~\cite{Exirifard:2007da,Tsujikawa:2007tg,Bauer:2008zj,Borunda:2008kf,Capozziello:2009nq,DeFelice:2010aj,Goenner:2010tr}.
Non-trivial properties in Palatini models have been investigated in
Refs.~\cite{Meng:2003sx,Sotiriou:2005cd,Li:2008bma,Li:2008fa}, and
finally the resolution of the Big Bang singularity has been proposed
in this context~\cite{Olmo:2009xy,Barragan:2010qb}, possibly in
connection with loop quantum gravity \cite{Olmo:2008nf}.

The paper is organised as follows, in Sec.~\ref{sec:fRQ-Palatini}
we briefly reproduce how to solve $f(R,Q)$ modified gravity models.
In Sec.~\ref{sec:CC-Filter} we present our model which filters out
the large CC. The results will be applied to cosmology in Sec.~\ref{sec:Cosmology}
discussing the radiation, matter and late-time de~Sitter era. Finally,
in Sec.~\ref{sec:Kottler} we show that the CC filter works also
for the Kottler (Schwarzschild-de~Sitter) solution describing a black
hole in the presence of a CC. We conclude in Sec.~\ref{sec:Conclusions}
and give an outlook to future developments.

In this work the speed of light~$c$ and the Planck constant~$\hbar$
are set to unity, the signature of the metric is $(-1,+1,+1,+1)$.

\section{$f(R,Q)$ modified gravity in the Palatini formalism}

\label{sec:fRQ-Palatini}The action of our setup is given by\begin{equation}
\mathcal{S}=\int d^{4}x\,\left[\sqrt{|g|}\frac{1}{2}f(R,Q)\right]+\mathcal{S}_{\text{mat}}[g_{ab,}\phi],\label{eq:P-fRQ-action}\end{equation}
where $g_{ab}$ is the {}``physical'' metric, on which the matter
fields~$\phi$ in $\mathcal{S}_{\text{mat}}$ propagate. The Ricci
scalar $R$ and the squared Ricci tensor~$Q$ depend on both the
metric and the connection~$\Gamma_{bc}^{a}$ while the Ricci tensor
$R_{ab}$ is defined only in terms of the latter:\begin{eqnarray}
R_{ab}[\Gamma] & = & \Gamma_{ab,e}^{e}-\Gamma_{eb,a}^{e}+\Gamma_{ab}^{e}\Gamma_{fe}^{f}-\Gamma_{af}^{e}\Gamma_{eb}^{f}\label{eq:P-RicciTensor}\\
R[g,\Gamma] & = & R_{\,\, a}^{a}=g^{ab}R_{ab}\label{eq:P-RicciScalar}\\
Q[g,\Gamma] & = & R^{ab}R_{ab}=g^{ac}g^{bd}R_{ab}R_{cd}.\label{eq:P-RicciQ}\end{eqnarray}
Note that in general these quantities are different from their metric
versions, and the absence of torsion implies that the connection is
symmetric. Moreover, we restrict our discussion to the case of a symmetric
Ricci tensor~$R_{ab}$ in the action~(\ref{eq:P-fRQ-action}). This
property is not automatic even for symmetric connections, which was
shown recently in Ref.~\cite{Vitagliano:2010pq}.

With these preliminaries the variation $2\,\delta\mathcal{S}/\delta g^{ab}=0$
of the action functional~$\mathcal{S}$ with respect to~$g_{ab}$
yields the modified Einstein equations\begin{equation}
f_{R}R_{m}^{\,\,\, n}+2f_{Q}R_{m}^{\,\,\, a}R_{a}^{\,\, n}-\frac{1}{2}\delta_{m}^{\,\,\, n}f=T_{m}^{\,\,\, n},\label{eq:P-EOM-1}\end{equation}
where the energy-momentum tensor~$T_{m}^{\,\,\, n}$ emerges from
the term $\mathcal{S}_{\text{mat}}$, and~$f_{R}$ and~$f_{Q}$
are partial derivatives of~$f$ with respect to the scalars~$R$
and~$Q$. Moreover, we obtain from $\delta\mathcal{S}/\delta\Gamma_{bc}^{a}=0$
the equation of motion~(EOM) for the Palatini connection,\begin{equation}
\nabla_{a}\left[\sqrt{|g|}(f_{R}g^{mn}+2f_{Q}R^{mn})\right]=0,\label{eq:P-EOM-2}\end{equation}
where~$\nabla_{a}$ denotes the covariant derivative in terms of
the yet unknown connection $\Gamma_{bc}^{a}$.

At this point we should remark that in the metric formalism the term~$Q$
yields EOMs with higher-order derivatives and problematic extra degrees
of freedom. Generally, it is difficult to avoid instabilities, e.g.\
of the Ostrogradski-type%
\footnote{In the metric approach this type of instability can be avoided in
gravity actions depending only on the Ricci scalar~$R$ and the Gau\ss-Bonnet
term~$G$. It would be interesting to compare both approaches in
the context of $f(R,G)$ models. However, $G$ contains the squared
Riemann tensor, and to our knowledge no method to solve the corresponding
Palatini EOMs has been found yet.%
}~\cite{Woodard:2006nt}. In contrast, the Palatini formalism provides
second-order EOMs for our scenario just as standard general relativity,
and problems from extra degrees of freedom do not occur.

In the following, we work along the lines of Ref.~\cite{Olmo:2009xy},
where a procedure for solving the Palatini EOMs is discussed. The
strategy is as follows: the formal solutions to both EOMs in~(\ref{eq:P-EOM-1})
and~(\ref{eq:P-EOM-2}) relate in an algebraic way the geometrical
scalars~$R$ and~$Q$ with the physical metric~$g_{ab}$ and the
energy-momentum tensor~$T_{m}^{\,\,\, n}$, respectively. Thus, one
can express~$R$ and~$Q$ in terms of the matter content alone by
solving this set of equations. Subsequently, the results are plugged
back into the formal solutions to obtain the explicit form of the
Palatini connection~$\Gamma$ and all the quantities derived from
it. Since the results will involve the metric~$g_{ab}$ and its derivatives,
it is possible to relate the cosmic expansion rate with the matter
energy density, for instance. In the rest of this section, we explain
the procedure for general~$f(R,Q)$ models.

First, let us write Eq.~(\ref{eq:P-EOM-1}) in matrix form by introducing
the $4\times4$-matrices $\hat{P}=R_{m}^{\,\, n}=R_{ma}g^{an}$ and
$\hat{T}=T_{m}^{\,\, n}$, whose entries are just the components of
the corresponding tensors components. With the identity matrix~$\hat{I}$
we obtain\begin{equation}
f_{R}\hat{P}+2f_{Q}(\hat{P})^{2}-\frac{1}{2}f\hat{I}=\hat{T},\label{eq:P-EOM-1-matrix}\end{equation}
and the trace of this equation reads\begin{equation}
f_{R}R+2Qf_{Q}-2f=T,\label{eq:P-EOM-1-trace}\end{equation}
where $R=\text{tr}(\hat{P})$, $Q=\text{tr}(\hat{P}^{2})$ and $T=T_{m}^{\,\, m}$.

For determining the Palatini connection~$\Gamma$ we introduce the
auxiliary metric~$h_{mn}$ and consider the equation\begin{equation}
\nabla_{a}[\Gamma]\left[\sqrt{|h|}h^{mn}\right]=0,\label{eq:P-EOM-2-h}\end{equation}
where $\nabla_{a}[\Gamma]$ is the covariant derivative in terms of~$\Gamma$.
Consequently, in order to solve this equation the connection~$\Gamma$
must be compatible with $h_{mn}$, i.e.\ it has to be the Levi-Civita
connection of $h_{mn}$,\begin{equation}
\Gamma_{bc}^{a}[h]=\frac{1}{2}h^{ad}(h_{dc,b}+h_{bd,c}-h_{bc,d}),\label{eq:LC-Con-h}\end{equation}
just as in general relativity. Hence, we convert Eq.~(\ref{eq:P-EOM-2})
into~(\ref{eq:P-EOM-2-h}) by defining~$h^{mn}$ in the following
way, \begin{equation}
\sqrt{|h|}\hat{h}^{-1}=\sqrt{|g|}\hat{g}^{-1}\hat{\Sigma}\,\,\,\,\text{with}\,\,\,\,\hat{\Sigma}=\left(f_{R}\hat{I}+2f_{Q}\hat{P}\right),\label{eq:P-SigmaDef}\end{equation}
where the metric~$h_{mn}$ (and analogously~$g_{mn}$) has been
written in matrix notation as $\hat{h}=h_{mn}$ with its inverse $\hat{h}^{-1}=h^{mn}$.
Calculating the determinant of both sides, we find $h^{2}h^{-1}=h=g\,\text{det \ensuremath{\hat{\Sigma}}}$,
which allows to eliminate $\sqrt{|h|}$ and finally yields \begin{equation}
h^{mn}=\hat{h}^{-1}=\frac{\hat{g}^{-1}\hat{\Sigma}}{\sqrt{|\text{det}\,\hat{\Sigma}|}},\,\,\, h_{mn}=\hat{h}=\sqrt{|\text{det}\,\hat{\Sigma}|}\,\hat{\Sigma}^{-1}\hat{g}.\label{eq:P-hmn-matrix}\end{equation}
Since the connection in~(\ref{eq:LC-Con-h}) solves Eq.~(\ref{eq:P-EOM-2-h}),
the formal solution of Eq.~(\ref{eq:P-EOM-2}) is also given by~$\Gamma_{bc}^{a}[h]$
in~(\ref{eq:LC-Con-h}) if $h_{mn}$ is defined as in~(\ref{eq:P-hmn-matrix}).

The remaining step is to find~$\hat{P}$ which requires an explicit
form for the energy-momentum tensor~$T_{m}^{\,\,\, n}$. Here, we
assume that the matter sector can be described by a perfect fluid,\begin{equation}
T_{m}^{\,\,\, n}=(\rho+p)u_{m}u^{n}+(p-\Lambda)\delta_{m}^{\,\, n},\label{eq:StressTensor}\end{equation}
where $\rho$ and $p$ are the energy density and pressure of (ordinary)
matter, including e.g.\ dust and incoherent radiation. And~$u_{m}$
denotes the corresponding $4$-velocity vector of the matter field.
$\Lambda$ represents the energy density corresponding to the cosmological
constant, and it contains all vacuum energy contributions. Thus we
require $p\neq-\rho$ without loss of generality. Next, let us write
the matrix expression~(\ref{eq:P-EOM-1-matrix}) in the following
way\begin{eqnarray}
(2f_{Q})^{2}\hat{M}^{2} & = & x^{2}\,\hat{I}+\mu\,\widehat{u_{m}u^{n}}\label{eq:P-EOM-1-MatrixForm}\\
\hat{M} & := & \hat{P}+\frac{1}{4}\frac{f_{R}}{f_{Q}}\,\hat{I}\label{eq:P-Matrix-M}\\
x^{2} & := & 2f_{Q}(p-\Lambda)+f_{Q}\, f+\frac{1}{4}f_{R}^{2}\label{eq:x2Def}\\
\mu & := & 2f_{Q}(\rho+p).\label{eq:P-mu}\end{eqnarray}
By explicit calculation one can check that\begin{equation}
c_{a}\cdot2f_{Q}\hat{M}=x\,\hat{I}+y\,\widehat{u_{m}u^{n}}\label{eq:P-EOM-1-sol}\end{equation}
with\begin{equation}
y:=\frac{-x+c_{b}\cdot\sqrt{x^{2}+\mu(u_{m}u^{m})}}{(u_{m}u^{m})}\label{eq:P-EOM-yDef}\end{equation}
is a solution to Eq.~(\ref{eq:P-EOM-1-MatrixForm}), which yields~$\hat{P}$.
The constants $c_{a,b}=\pm1$ and the sign convention%
\footnote{Changing the sign of $x$ in Eqs.~(\ref{eq:P-EOM-1-sol}) and~(\ref{eq:P-EOM-yDef})
by $x\rightarrow-x$ corresponds to $c_{a}\rightarrow-c_{a}$, and
therefore it does not yield a new solution. Here, we use the convention
$\sqrt{x^{2}+\mu(u_{m}u^{m})}=x\sqrt{1+\mu(u_{m}u^{m})/x^{2}}$, and
the second possibility~$-x\sqrt{1+\cdots}$ would just mean $c_{b}\rightarrow-c_{b}$.%
} for~$x=\sqrt{x^{2}}$ will be fixed later by consistency considerations.

Now, we have to determine the scalars $R$ and $Q$, which follow
from the trace equation~(\ref{eq:P-EOM-1-trace}) and the trace of~(\ref{eq:P-EOM-1-sol}),\begin{equation}
c_{a}(2f_{Q}R+2f_{R})=4x-y,\label{eq:P-EOM-1-sol-trace}\end{equation}
where $u_{m}u^{m}=-1$ will be used from here on. In the last equation
we eliminate all roots by squaring twice, which results to\begin{equation}
\left(\left(2f_{Q}R+2f_{R}\right)^{2}+8x^{2}+\mu\right)^{2}=36x^{2}\left(2f_{Q}R+2f_{R}\right)^{2},\label{eq:P-BigEq}\end{equation}
where~$c_{a,b}=\pm1$ and odd powers of~$x$ have dropped out. Solving
this algebraic equation together with~(\ref{eq:P-EOM-1-trace}) is
the tough part of the Palatini formalism. Once this task is achieved,
$R$,~$Q$ and $\hat{P}=R_{m}^{\,\, n}$ in~(\ref{eq:P-EOM-1-sol})
are given as functions of $\rho,p,\Lambda$ only, and can be used
to calculate the connection~$\Gamma$. For this purpose, let us write
the matrix~$\hat{\Sigma}$ in~(\ref{eq:P-SigmaDef}) as\begin{eqnarray}
\hat{\Sigma} & = & 2f_{Q}\hat{M}+\frac{1}{2}f_{R}\hat{I}=L_{1}\,\hat{I}+L_{2}\,\widehat{u_{m}u^{n}}=L_{1}\left(\hat{I}+\frac{L_{2}}{L_{1}}\,\widehat{u_{m}u^{n}}\right),\label{eq:SigmaSol}\\
 &  & \text{with}\,\,\,\, L_{1}:=c_{a}x+\frac{1}{2}f_{R},\,\,\,\,\,\,\,\, L_{2}:=c_{a}y.\label{eq:L1L2Def}\end{eqnarray}
Using $\text{det}(\hat{I}+\widehat{a_{m}b^{n}})=1+b^{m}a_{m}$ we
find $\text{det}(\hat{\Sigma})=L_{1}^{3}(L_{1}-L_{2})$, and the inverse
matrix reads\[
\hat{\Sigma}^{-1}=\frac{1}{L_{1}}\,\hat{I}-\frac{L_{2}}{L_{1}(L_{1}-L_{2})}\,\widehat{u_{m}u^{n}}.\]
Finally, we have all ingredients to write down the explicit form of
the auxiliary metric%
\footnote{The relation in Eq.~(\ref{eq:h_mn}) is called a disformal transformation~\cite{Bekenstein:1992pj},
which is used e.g.\ in MOND theories~\cite{Bekenstein:2004ne} and
scalar field models for dark energy~\cite{Zumalacarregui:2010wj}.%
} from Eq.~(\ref{eq:P-hmn-matrix})\begin{eqnarray}
h_{mn} & = & \Omega\left(g_{mn}-\frac{L_{2}}{L_{1}-L_{2}}u_{m}u_{n}\right)\label{eq:h_mn}\\
h^{mn} & = & \Omega^{-1}\left(g^{mn}+\frac{L_{2}}{L_{1}}u^{m}u^{n}\right),\end{eqnarray}
where \begin{equation}
\Omega:=\frac{\sqrt{|\text{det}\hat{\Sigma}|}}{L_{1}}=\frac{\sqrt{|L_{1}^{3}(L_{1}-L_{2})|}}{L_{1}}.\label{eq:OmegaDef}\end{equation}
Once~$h_{mn}$ is known, the connection~$\Gamma_{bc}^{a}$ follows
directly from Eq.~(\ref{eq:LC-Con-h}), and subsequently the Ricci
tensor~(\ref{eq:P-RicciTensor}) and the scalars in Eqs.~(\ref{eq:P-RicciScalar})
and~(\ref{eq:P-RicciQ}) can be calculated.

\section{Relaxing the CC with a filter}

\label{sec:CC-Filter}In this section we study a modified gravity
model to relax the CC in the Palatini $f(R,Q)$ framework. Motivated
by earlier work in the metric formalism~\cite{Bauer:2009ea,Bauer:2010wj}
we consider the following ansatz for the gravity action\begin{equation}
f(R,Q)=\kappa R+z\,\,\,\text{with}\,\,\, z:=\beta\,\frac{R^{n}}{B^{m}},\,\,\, B=R^{2}-Q,\label{eq:fRQ-ansatz}\end{equation}
where~$\kappa$ and $\beta$ are non-zero constant parameters and
$n$ and~$m$ positive numbers. In the metric formalism model~\cite{Bauer:2009ea}
the structure of the function~$B$ was enforcing the universe to
expand like a matter dominated cosmos even when the matter energy
density~$\rho_{m}$ was much smaller in magnitude than the vacuum
energy density~$\Lambda$. However, in the Palatini framework the
function~$B$ is not known in the beginning, and it is necessary
to investigate under which circumstances a relaxed cosmological expansion
behaviour can be obtained. Moreover, we will see in the following
that~$z$ is not a correction to the Einstein-Hilbert term but a
crucial part of the action functional~$f$. Therefore, one should
refrain from considering the limit~$z\rightarrow0$.

From Eq.~(\ref{eq:fRQ-ansatz}) we find\begin{equation}
f_{R}=\frac{\kappa R+nz}{R}-2f_{Q}R,\,\,\,\, f_{Q}=m\frac{z}{B},\label{eq:fRfQ}\end{equation}
and the trace of the stress tensor~(\ref{eq:StressTensor}) reads\begin{equation}
T=-4\Lambda+3p-\rho.\label{eq:T-stress}\end{equation}
As a result, Eq.~(\ref{eq:P-EOM-1-trace}) provides the first equation
for finding~$R$ and~$B$ (or $Q$),\begin{equation}
\gamma z=\kappa R-4\Lambda+3p-\rho,\,\,\,\,\,\gamma:=(n-2-2m),\label{eq:z}\end{equation}
while the second one is given in Eq.~(\ref{eq:P-BigEq}), explicitly\begin{eqnarray}
0 & = & f_{Q}^{3}R^{2}\, S_{3}+f_{Q}^{2}\, S_{2}+f_{Q}R^{-2}\, S_{1}\label{eq:SeqZero}\\
S_{3} & := & 72\left[\kappa R+\rho+p+\frac{1}{3}z(2+2n+\gamma)\right]\label{eq:S3}\\
S_{2} & := & 4\left[z^{2}\left(-17n^{2}+4n(2+\gamma)+4(2+\gamma)^{2}\right)\right.\nonumber \\
 & + & 12z(\rho+p)(2-n+\gamma)+6z\kappa R(4-5n+2\gamma)\nonumber \\
 & + & \left.9(\rho+p)^{2}-9(\kappa R)^{2}\right]\label{eq:S2}\\
S_{1} & := & 24\left[z(\kappa R+nz)^{2}(n-2-\gamma)\right].\label{eq:S1}\end{eqnarray}
Apparently, this set of equations is quite complicated, thus we will
solve it approximately. First, we consider in this work only the epochs
when the large CC~$\Lambda$ dominates over all other energy sources.
Then Eq.~(\ref{eq:z}) implies the relation~$z=\beta R^{n}/B^{m}=\mathcal{O}(\Lambda)$
suggesting that~$B/R^{2}$ and $f_{Q}^{-1}=B/(mz)$ are relatively
small quantities, which may be used as expansion parameters in Eq.~(\ref{eq:SeqZero}).
At this point we cannot prove this suggestion quantitatively because~$R$
and~$B$ are not known yet. However, it will be confirmed later by
Eqs.~(\ref{eq:BR2-early}) and~(\ref{eq:BR2-late}). As a result
of assuming that $f_{Q}^{-1}$ is sufficiently small, the term~$S_{3}$
proportional to $f_{Q}^{3}$ in~(\ref{eq:SeqZero}) will be the most
important one, and a good zero-order solution can be found by neglecting
the other terms $f_{Q}^{2}\, S_{2}+f_{Q}R^{-2}\, S_{1}$ and solving
only~$S_{3}=0$.

Our goal is a relaxed universe, i.e.\ one which is not dominated
by the large CC term, and it can be realised by requiring that the
Ricci scalar~$R$ following from $S_{3}=0$ is free from large~$\mathcal{O}(z)$
contributions. This happens when the parameter~$n$ is restricted
by $2+2n+\gamma=3n-2m=0$, which eliminates the $\mathcal{O}(z)$
term in Eq.~(\ref{eq:S3}). We will apply this condition from now
on, hence Eq.~(\ref{eq:SeqZero}) can be written as\begin{eqnarray}
0 & = & \kappa R+r\label{eq:SeqZeroRelax}\\
 & - & \frac{2}{9}mz\left(\frac{B}{R^{2}}\right)\left[1+\frac{3}{mz}\left(3\kappa R+2r\right)+\frac{9}{4(mz)^{2}}\left((\kappa R)^{2}-r^{2}\right)\right]\nonumber \\
 & + & \frac{8}{27}mz\left(\frac{B}{R^{2}}\right)^{2}\left(1+\frac{3\kappa R}{2(mz)}\right)^{2}\,\,\,\,\text{with}\,\,\,\, r:=\rho+p.\nonumber \end{eqnarray}
From the first line in the last equation one clearly observes that
CC terms with EOS~$p=-\rho$ do not contribute to the Ricci scalar
at leading order. In other words, the CC is filtered out from~$r=\rho+p$,
which describes the matter sector. Note that~$z=\mathcal{O}(\Lambda)$
still appears in suppressed correction terms.

In the following, we will solve Eq.~(\ref{eq:SeqZeroRelax}) in situations
relevant for cosmology. A first order correction to $S_{3}=0$ can
be found by keeping only the term $z(B/R^{2})$ from the second line
in~(\ref{eq:SeqZeroRelax}), which leads to\begin{equation}
\kappa R+r=\frac{2}{9}mz\left(\frac{B}{R^{2}}\right)=\frac{2}{9}mz\left(\frac{\beta}{z}\right)^{\frac{1}{m}}R^{-\frac{4}{3}},\end{equation}
where we used $z=\beta R^{n}/B^{m}$ from Eq.~(\ref{eq:fRQ-ansatz})
with $3n=2m$ in the last step. Thus, we obtain a $7^{\text{th}}$-order
polynomial equation in~$R$,\begin{equation}
\left(\kappa R+r\right)^{3}(\kappa R)^{4}=L_{3}:=\kappa^{4}\left(\frac{2}{9}mz\left(\frac{\beta}{z}\right)^{\frac{1}{m}}\right)^{3},\label{eq:kRpolyEqu}\end{equation}
which clearly shows that~$R$ is a function of~$\rho$ and~$p$
only. Finally, we find the approximate solution\begin{equation}
\kappa R=-r+D+\mathcal{O}\left(\frac{D^{2}}{r}\right)\,\,\,\text{with}\,\,\, D:=\left(\frac{L_{3}}{r^{4}}\right)^{\frac{1}{3}},\label{eq:kR-early}\end{equation}
when the matter-related quantity~$r$ lies in the range $|D|\ll r\ll|\Lambda|$,
which we refer to as the early-time limit or the limit of large energy
density in~$r$. Moreover, $\kappa R<0$ because $r>0$ for ordinary
matter, and~$\kappa R$ will remain negative for decreasing~$\rho$
because $\kappa R\rightarrow0$ is not a solution of Eq.~(\ref{eq:kRpolyEqu}).

\begin{figure}
\begin{centering}
\includegraphics[width=1\columnwidth]{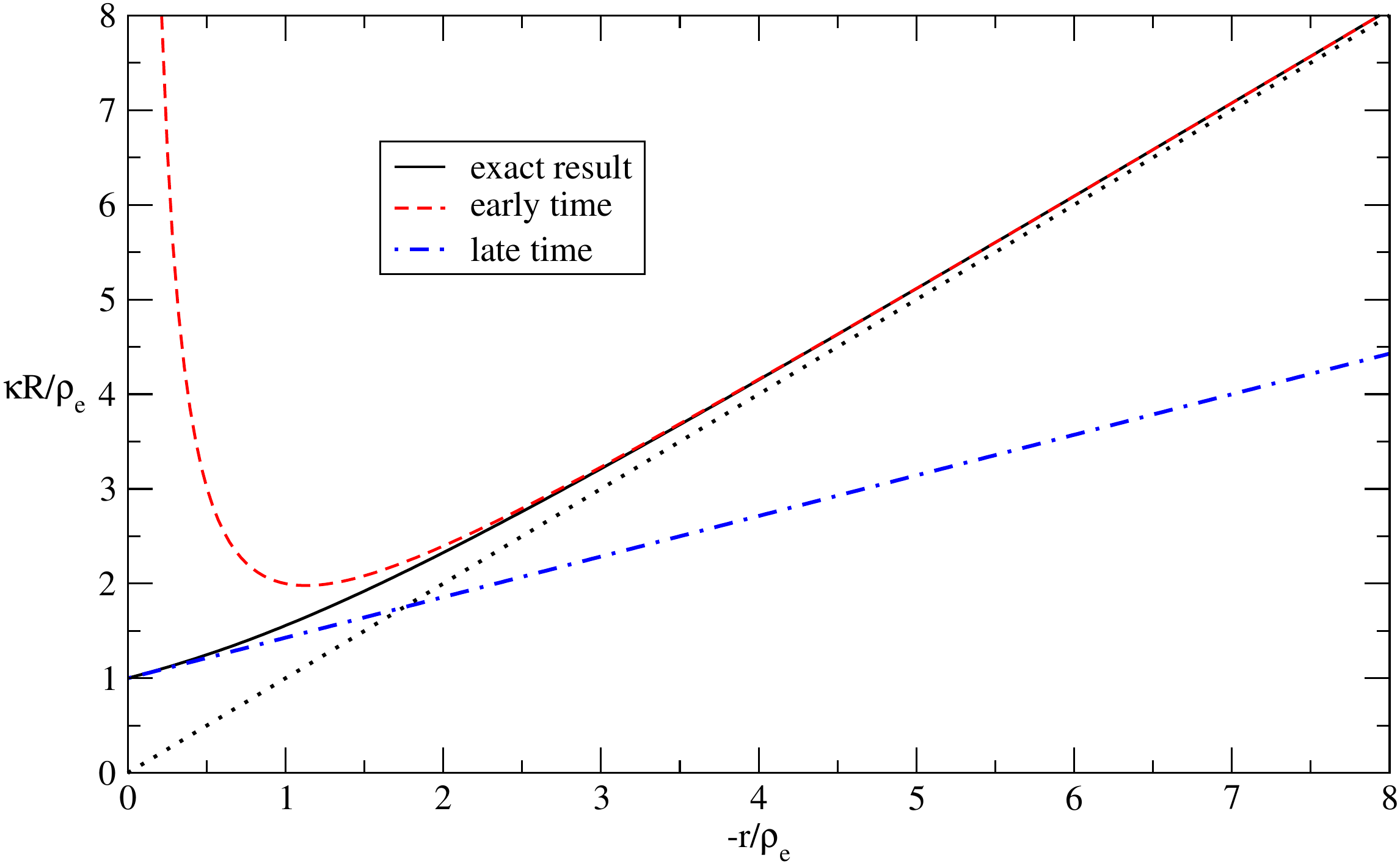}
\par\end{centering}

\caption{Numerical solutions for~$\kappa R<0$ as a function of~$r=\rho$
with $p=0$. The plot shows the exact result in Eq.~(\ref{eq:SeqZeroRelax})
(black solid curve) and the approximations at early times (red dashed)
from Eq.~(\ref{eq:kR-early}) as well as at late times (blue dashed-dotted)
from Eq.~(\ref{eq:kR-late}), respectively. In this example with~$m=3$
the parameter~$\beta<0$ has been chosen such that $-|\Lambda|\approx358\,\rho_{e}$
only for numerical reasons. A more realistic ratio of~$|\Lambda/\rho_{e}|$
would be much larger, however the results would not differ qualitatively,
which is true also for other values of the EOS~$p/\rho$ as long
as~$r\ll|\Lambda|$. In any case,~$\kappa R$ approaches~$-r$
(dotted diagonal line) in the region~$-\rho_{e}\ll r\ll|\Lambda|$.\label{fig:kRofr-m3}}

\end{figure}

In the opposite limit, $r\rightarrow0$, which is called the late-time
limit in the following, we obtain from Eq.~(\ref{eq:kRpolyEqu})\begin{equation}
\kappa R=\rho_{e}-\frac{3}{7}r+\mathcal{O}\left(\frac{r^{2}}{\rho_{e}}\right)\,\,\,\text{with}\,\,\,\rho_{e}:=L_{3}^{\frac{1}{7}},\label{eq:kR-late}\end{equation}
indicating that $\kappa R$ approaches the negative constant~$\rho_{e}$
for vanishing matter. For a given value of~$\Lambda$ or $z$ the
parameter~$\beta$ must be chosen adequately for~$L_{3}=\rho_{e}^{7}<0$.
In Sec.~\ref{sub:Cosmo-Late} we will show that~$(-\rho_{e})$ is
close to the critical energy density in the asymptotic future, which
corresponds to the tiny observed value of the effective CC in Eq.~(\ref{eq:Late-He-rhoe}).
The parameter dependence of the solution set of the full equation~(\ref{eq:SeqZeroRelax})
might further constrain~$\beta$, however this has to be determined
numerically. In Fig.~\ref{fig:kRofr-m3} we show an example for~$m=3$
and~$\rho_{e}<0$, which nicely demonstrates the validity of our
approximations. Note that not all values of~$m$ might allow physically
reasonable solutions. We will discuss some examples for~$\beta$
at the end of Sec.~\ref{sub:Cosmo-Late}. However, in this work we
concentrate on analytical results, and the complete parameter dependence
as well as the case~$L_{3}>0$ will be investigated elsewhere.

Via $z=\beta(R^{2/3}/B)^{m}$ from Eq.~(\ref{eq:fRQ-ansatz}) it
is straightforward to obtain~$B$ and $B/R^{2}$ from the approximate
solution of $R$. In the following we denote subdominant corrections
by~$\varepsilon\ll1$. Accordingly, at early times Eq.~(\ref{eq:kR-early})
yields\begin{eqnarray}
B & = & \sqrt{\frac{L_{3}}{D}}\left(\frac{2}{9}mz\right)^{-1}(-\kappa)^{-2}\left(1-\frac{2}{3}\frac{D}{r}+\mathcal{O}(\varepsilon^{2})\right),\label{eq:B-early}\\
\frac{B}{R^{2}} & = & \frac{D}{\left(\frac{2}{9}mz\right)}\left(1-\frac{D}{r}+\mathcal{O}(\varepsilon^{2})\right),\label{eq:BR2-early}\end{eqnarray}
whereas from Eq.~(\ref{eq:kR-late}) we obtain the corresponding
late-time results\begin{eqnarray}
B & = & \rho_{e}^{3}\left(\frac{2}{9}mz\right)^{-1}(-\kappa)^{-2}\left(1-\frac{2}{7}\frac{r}{\rho_{e}}+\mathcal{O}(\varepsilon^{2})\right),\label{eq:B-late}\\
\frac{B}{R^{2}} & = & \rho_{e}\left(\frac{2}{9}mz\right)^{-1}\left(1-\frac{4}{7}\frac{r}{\rho_{e}}+\mathcal{O}(\varepsilon^{2})\right).\label{eq:BR2-late}\end{eqnarray}
These results confirm that~$B/R^{2}$ is sufficiently small to justify
that we neglected some terms in Eq.~(\ref{eq:SeqZeroRelax}), hence,
we have found consistent solutions for~$R$ and~$B$. More support
for the validity of the approximations is provided by the numerical
solution of the complete Eq.~(\ref{eq:SeqZeroRelax}) in Fig.~\ref{fig:kRofr-m3}.

With~$R$ and $B$ as functions of only the matter-related term~$r$,
it is possible to calculate~$L_{1,2}$ in~(\ref{eq:L1L2Def}). First,
we have to define the square root of $x^{2}$ in~(\ref{eq:P-EOM-1-sol}).
Since we work in the limit $|R^{2}/B|\gg1$, which implies $|f_{Q}R|=|mzR/B|\gg z/R$,
we identify $f_{Q}R$ to be the dominant term in Eqs.~(\ref{eq:fRfQ})
and~(\ref{eq:x2Def}). Thus $f_{R}\approx-2f_{Q}R$ and $x^{2}\approx\frac{1}{4}f_{R}^{2}\approx(f_{Q}R)^{2}$,
and we choose the convention~$x=f_{Q}R\sqrt{1+\cdots}$, where~$\cdots$
denotes the remaining terms in Eq.~(\ref{eq:x2Def}) divided by~$(f_{Q}R)^{2}$.
Next, we plug $x\approx-\frac{1}{2}f_{R}\approx f_{Q}R$ into the
trace equation~(\ref{eq:P-EOM-1-sol-trace}) \begin{equation}
c_{a}(2f_{Q}R+2f_{R})=3x+c_{b}\cdot x\sqrt{1-(2f_{Q}r)x^{-2}},\end{equation}
which gives in leading order $c_{a}(-2)f_{Q}R=f_{Q}R(3+c_{b})$. Obviously,
$c_{a}=c_{b}=-1$ is the only solution%
\footnote{Note that our scenario with $c_{b}=-1$ relies from the very beginning
on the existence of matter, because otherwise the vector~$u_{m}$
in Eq.~(\ref{eq:P-EOM-1-MatrixForm}) were absent and~$c_{b}=+1$
would be required in Eq.~(\ref{eq:P-EOM-yDef}). However, matter
is a fact of reality, which singles out~$c_{b}=-1$ in our model.
Consequently, Eq.~(\ref{eq:kR-late}) is the correct solution in
the~$r\rightarrow0$ limit.%
} in the large $|f_{Q}R|$ limit, and we find\begin{eqnarray}
L_{1} & = & -x+\frac{1}{2}f_{R}\nonumber \\
 & = & -2f_{Q}R+R^{-1}\left(mz+\frac{3}{4}\kappa R-\frac{1}{4}r\right)+\frac{(mz)^{2}}{6f_{Q}R^{3}}\left(1+\varepsilon\right),\label{eq:L1-Sol}\\
L_{2} & = & -x-x\sqrt{1-(2f_{Q}r)x^{-2}}\nonumber \\
 & = & -2f_{Q}R+R^{-1}\left(\frac{4}{3}mz+\frac{1}{2}(\kappa R+r)\right)+\frac{(mz)^{2}}{3f_{Q}R^{3}}\left(1+\varepsilon\right),\label{eq:L2-Sol}\end{eqnarray}
where we used a series expansion of the root in\begin{eqnarray}
f_{R} & = & -2f_{Q}R+R^{-1}\left(\kappa R+\frac{2}{3}mz\right),\\
x & = & f_{Q}R\sqrt{1-\frac{8mz+3\kappa R-3r}{6f_{Q}R^{2}}+\frac{(2mz+3\kappa R)^{2}}{36f_{Q}^{2}R^{4}}}.\end{eqnarray}
Above and from here on~$\varepsilon$ denotes small terms of the
order~$\kappa R/z$, $r/z$ or~$B/R^{2}$. Moreover, $\Lambda$
has been eliminated in favour of\begin{equation}
z=-\frac{3}{6+4m}\left(\kappa R-4\Lambda+3p-\rho\right).\end{equation}
Finally, we write down the series expansion of\begin{equation}
L_{1}-L_{2}=-\frac{1}{3}\frac{mz}{R}\left(1+\frac{1}{2}\frac{B}{R^{2}}\left(1+\varepsilon\right)\right),\label{eq:L1-L2-Sol}\end{equation}
which appears in the auxiliary metric~$h_{ab}$. Now, we have all
ingredients available for discussing solutions to the Palatini field
equations in the next sections.

\section{Cosmology}

\label{sec:Cosmology}For investigating the cosmological evolution
the physical metric is of the spatially flat Friedmann-Lema\^itre-Robertson-Walker~(FLRW)
type with $g_{00}=-1$ and $g_{ii}=a^{2}(t)$ in Cartesian coordinates,
where~$a(t)$ is the scale factor as a function of cosmological time~$t$.
The matter component is at rest ($u^{m}=-\delta_{0}^{m}$) in these
coordinates. Accordingly, the auxiliary metric~$h_{ab}$ in Eq.~(\ref{eq:h_mn})
is completely defined by the diagonal elements $h_{ii}=\Omega g_{ii}$
and $h_{00}=\Omega(g_{00}-L_{2}/(L_{1}-L_{2}))$, which follow explicitly
from Eqs.~(\ref{eq:L1-Sol}),~(\ref{eq:L2-Sol}) and~(\ref{eq:L1-L2-Sol}),
\begin{eqnarray}
\Omega & = & \frac{\sqrt{|L_{1}(L_{1}-L_{2})|}}{L_{1}}=\text{sgn}(L_{1})\sqrt{|\frac{2}{3}(mz)^{2}B^{-1}|}\left(1+\varepsilon^{2}\right),\label{eq:Om-Cosmo}\\
h_{00} & = & \Omega\frac{-L_{1}}{L_{1}-L_{2}}=\Omega\,\left(-6\frac{R^{2}}{B}\right)\left(1-\frac{B}{R^{2}}\left(1+\varepsilon\right)\right).\label{eq:h00-Cosmo}\end{eqnarray}
Note that if~$B$ and~$L_{1}$ are positive, the metric~$h_{ab}$
has the same signature as~$g_{ab}$. Moreover, in the following~$z$
will be treated as a time-independent constant proportional to the~CC~$\Lambda$
since the corrections can be subsumed in the $\varepsilon\sim\mathcal{O}(\kappa R/z)$
terms. At this point we are ready to apply the results for~$B$ and~$R^{2}/B$
that we found in Eqs.~(\ref{eq:B-early}),~(\ref{eq:BR2-early}),~(\ref{eq:B-late})
and~(\ref{eq:BR2-late}), respectively.

\subsection{Early universe}

\label{sub:Cosmo-Early}We begin with the epoch, where the matter
energy density~$\rho$ (including dust and radiation) is well below
the large CC~$\Lambda\sim z$ in magnitude but above the asymptotic
future energy density $|\rho_{e}|$. Therefore, our discussion will
be valid for most parts of the radiation and matter eras. According
to~(\ref{eq:Om-Cosmo}) and~(\ref{eq:h00-Cosmo}) we find\begin{eqnarray}
\Omega & = & c_{1}\left(\frac{D}{L_{3}}\right)^{\frac{1}{4}}\left(1+e\, e_{1}\frac{D}{3r}+\mathcal{O}(e^{2})\right),\,\,\,\, c_{1}=\text{const.}\label{eq:Om-early}\\
h_{00} & = & c_{1}\left(-\frac{4}{3}mz/e\right)L_{3}^{-\frac{1}{4}}D^{-\frac{3}{4}}\left(1-e\, e_{2}\frac{D}{r}+\mathcal{O}(e^{2})\right),\label{eq:h00-early}\end{eqnarray}
where~$r=\rho+p$ and~$D=(L_{3}/r^{4})^{(1/3)}$ was introduced
earlier in Eq.~(\ref{eq:kR-early}). The sign under the root in the
constant~$c_{1}=\sqrt{\pm3(\frac{2}{9}mz)^{3}(-\kappa)^{2}}$ may
be chosen such that consistency with Eq.~(\ref{eq:Om-Cosmo}) is
obtained. However, since~$c_{1}$ drops out from the connection~$\Gamma(h)$
in Eq.~(\ref{eq:LC-Con-h}) we skip this question for the moment.
In addition, we have introduced above the purely technical parameter~$e=1$
which counts the powers of the small quantity~$|D|\ll r$, whereas
$e_{1,2}=1$ just denote first-order correction terms.

According to Eq.~(\ref{eq:LC-Con-h}) the non-zero components of
the (symmetric) connection read (with~$i=1,2,3$)\begin{eqnarray}
\Gamma_{00}^{0} & = & \frac{1}{2}\frac{\dot{r}}{r}+e\, e_{2}\frac{7}{6}\frac{\dot{r}}{r}\frac{D}{r}+\mathcal{O}(e^{2}),\\
\Gamma_{ii}^{0} & = & e\frac{a^{2}}{8mz}D\left(6\frac{\dot{a}}{a}-\frac{\dot{r}}{r}\right)+\mathcal{O}(e^{2}),\\
\Gamma_{i0}^{i} & = & \frac{1}{6}\left(6\frac{\dot{a}}{a}-\frac{\dot{r}}{r}\right)-e\, e_{1}\frac{7}{18}\frac{\dot{r}}{r}\frac{D}{r}+\mathcal{O}(e^{2}).\end{eqnarray}
For a simpler presentation we assume from now on a constant matter
EOS~$\omega=p/\rho>-1$, which yields a simple expansion law for
our matter component via its conservation equation%
\footnote{Since the matter action~$\mathcal{S}_{\text{mat}}$ in~(\ref{eq:P-fRQ-action})
does not involve the Palatini connection, the covariant derivative~$\nabla_{n}=\nabla_{n}[g]$
in the matter conservation equation~$\nabla_{n}T_{m}^{\,\,\, n}=0$
contains only the Christoffel symbols of the physical metric~$g_{ab}$.%
}\begin{equation}
\dot{\rho}+3\frac{\dot{a}}{a}\rho(1+\omega)=0\,\,\,\Leftrightarrow\,\,\,\rho\propto a^{-3(1+\omega)}.\label{eq:rhoOfa}\end{equation}
Hence, from Eq.~(\ref{eq:P-RicciTensor}) we obtain for the non-zero
components $R_{0}:=R_{0}^{\,\,0}$ and $R_{i}:=R_{1}^{\,\,1}=R_{2}^{\,\,2}=R_{3}^{\,\,3}$
of the diagonal Ricci tensor~$R_{a}^{\,\, b}=g^{cb}R_{ac}$ the following
results,\begin{eqnarray}
R_{0} & = & g^{00}R_{00}=\frac{3}{2}(\omega+3)\left(2(\omega+1)\left(\frac{\dot{a}}{a}\right)^{2}+\frac{\ddot{a}}{a}\right)\nonumber \\
 & + & e\frac{7}{4}\frac{D}{\rho_{m}}\left[\left(\frac{\dot{a}}{a}\right)^{2}\left(e_{1}(19\omega+21)+e_{2}3(\omega+3)\right)+e_{2}2\frac{\ddot{a}}{a}\right]+\mathcal{O}(e^{2}),\label{eq:R0ofa-early}\\
R_{i} & = & g^{11}R_{11}=e\frac{3D}{8mz}(\omega+3)\left[\left(\frac{\dot{a}}{a}\right)^{2}(3\omega+5)+\frac{\ddot{a}}{a}\right]+\mathcal{O}(e^{2}).\label{eq:Riofa-early}\end{eqnarray}
Remember that the Palatini connection~$\Gamma_{bc}^{a}$ given above
and all derived quantities like~$R$ or~$R_{ab}$ are different
from the metric versions, and they should not be compared with them.
Observables are related to the scale factor~$a$ in the physical
metric~$g_{ab}$. Due to~$R_{i}=\mathcal{O}(e)$ the leading component
of the Ricci scalar $R=R_{0}+3R_{i}$ is just the $\mathcal{O}(e^{0})$-term
in~$R_{0}$:\begin{equation}
R=\frac{3}{2}(\omega+3)\left(2(\omega+1)\left(\frac{\dot{a}}{a}\right)^{2}+\frac{\ddot{a}}{a}\right)+\mathcal{O}(e).\label{eq:R-early-0order}\end{equation}
Comparing this result with $\kappa R=-\rho(1+\omega)+\mathcal{O}(e)$
from Eq.~(\ref{eq:kR-early}), we quickly deduce the leading scale
factor behaviour. The situation is similar to general relativity because
the power-law ansatz~$a(t)\propto t^{s}$, $s=\text{const.}$ for
the scale factor represents a reasonable solution. Equipped with the
corresponding Hubble rate~$H=s/t$ and $\ddot{a}/a=s(s-1)/t^{2}$,
Eq.~(\ref{eq:R-early-0order}) yields $R\propto t^{-2}$ at zero
order, and $\rho\propto a^{-3(1+\omega)}\propto t^{-3s(1+\omega)}$
will be proportional to~$R$ if\begin{equation}
s=\frac{2}{3(\omega+1)}\,\,\,\mbox{in}\,\,\, a(t)\propto t^{s}.\end{equation}
Hence, we find from Eq.~(\ref{eq:R-early-0order})\begin{equation}
H=\frac{2}{3(\omega+1)t},\,\,\,\frac{\ddot{a}}{a}=H^{2}\frac{1+3\omega}{2},\,\,\, R=\frac{3}{4}(\omega+3)^{2}H^{2}+\mathcal{O}(e)=(-\kappa)^{-1}\rho(1+\omega),\end{equation}
implying the modified Friedmann equation\begin{equation}
(-\kappa)H^{2}=\frac{4}{3}\frac{\omega+1}{(\omega+3)^{2}}\,\rho,\label{eq:Friedmann-Early}\end{equation}
which is the main result of this section. Accordingly, a dust dominated
universe with EOS~$\omega=0$ leads to \begin{equation}
(-\kappa)H^{2}=\frac{4}{27}\rho,\,\,\,\,\,(\text{dust})\label{eq:Friedmann-dust}\end{equation}
which is not very different from the radiation dominated cosmos with
$\omega=\frac{1}{3}$. In the latter case, the modified Friedmann
equation reads \begin{equation}
(-\kappa)H^{2}=\frac{4}{25}\rho,\,\,\,\,\,(\text{radiation})\label{eq:Friedmann-rad}\end{equation}
where~$\rho$ is the radiation energy density. Obviously, dust matter
and radiation influence the cosmic expansion in almost the same way
as in standard general relativity if we choose the parameter~$\kappa=-\frac{4}{27}\cdot3/(8\pi G_{N})$
with Newton's constant~$G_{N}$. However, comparing Eq.~(\ref{eq:Friedmann-dust})
with~(\ref{eq:Friedmann-rad}) indicates an increased expansion rate
in the radiation era, which can be expressed by a higher effective
Newton constant~$G_{\text{rad}}=\frac{27}{25}G_{N}=1,08\, G_{N}$.
This $8\%$ difference seems to be well within current bounds on the
variation of~$G_{N}$. For instance, in the context of Big Bang nucleosynthesis
the bound~$G_{\text{rad}}/G_{N}=1,10\pm0,07$ was given recently
in Ref.~\cite{Steigman:2010zz}, which can be related also to constraints
on (additional) relativistic degrees of freedom~\cite{Izotov:2010ca,GonzalezGarcia:2010un,Hamann:2010bk}.
Accordingly, the small difference in the expansion rates above is
not in conflict with recent observations.

As a more technical point we remark that $\kappa=+1/(8\pi G_{N})$
is positive in general relativity, however, in our setup the negative
term~$\kappa R<0$ is only one part in the action~(\ref{eq:fRQ-ansatz}),
which is completed by the crucial term~$z$. Consequently, $\kappa<0$
does not correspond to a negative Newton constant in general relativity.
Another interesting feature of Eq.~(\ref{eq:Friedmann-Early}) is
that even if we had allowed additional vacuum energy ($\omega=-1$)
contributions in~$\rho$, it would not influence the expansion rate~$H$
because of the vanishing right-hand side of Eq.~(\ref{eq:Friedmann-Early}).
An example for this are finite shifts~$|\delta\Lambda|\ll|\Lambda|$
in the vacuum energy density emerging from phase transitions.

Comparing the Palatini setup with the CC relaxation models in the
LXCDM framework~\cite{Bauer:2009ke,Bauer:2009jk} and in metric $f(R,G)$~\cite{Bauer:2009ea,Bauer:2010wj}
modified gravity, we find another advantage. Here, it is not necessary
to include a parameter which controls the transition from radiation
to (dust) matter domination. It simply happens when the energy density
$\rho_{\text{dust}}\propto a^{-3}$ of dust matter overtakes the radiation
density $\rho_{\text{radiation}}\propto a^{-4}$ as a result of the
growing scale factor~$a$. Therefore, the relation of the energy
content with the expansion of the universe is very similar to general
relativity with a negligible CC.

For completeness we list all results including a first order correction
in the scale factor\begin{equation}
a(t)=\left(\frac{t}{t_{0}}\right)^{\frac{2}{3(\omega+1)}}\left(1+e_{a}\, e\frac{D}{\rho(1+\omega)}\right),\end{equation}
with the constant~$e_{a}$ to be determined below. The components
of the diagonal Ricci tensor read \begin{eqnarray}
R_{0} & = & \frac{(\omega+3)^{2}}{3t^{2}(1+\omega)^{2}}\left(1+e\frac{D}{\rho}\left[\frac{7[4e_{1}(5+4\omega)+3(3+\omega)(e_{2}+e_{a}(23+19\omega))]}{3(3+\omega)^{2}}\right]+\mathcal{O}(e^{2})\right),\nonumber \\
R_{i} & = & e\frac{D(\omega+3)^{2}}{4mzt^{2}(1+\omega)^{2}}+\mathcal{O}(e^{2}),\end{eqnarray}
yielding the Ricci scalar~$R=R_{0}+3R_{i}$ and the squared Ricci
tensor~$Q$,\begin{eqnarray}
R & = & R_{0}+\frac{(\omega+3)^{2}}{3t^{2}(1+\omega)^{2}}\left(1+e\frac{9}{4}\frac{D}{mz}+\mathcal{O}(e^{2})\right)\label{eq:Roft-early-full}\\
Q & = & R_{a}^{\,\, b}R_{b}^{\,\, a}=(R_{0})^{2}+3(R_{i})^{2}=(R_{0})^{2}+\mathcal{O}(e^{2}).\end{eqnarray}
Thus, the function~$B$ in the denominator of~$z$ in Eq.~(\ref{eq:fRQ-ansatz})
is given by\begin{equation}
B=R^{2}-Q=6R\cdot R_{i}-12(R_{i})^{2}=e\frac{D(3+\omega)^{4}}{mz(2t^{4})(1+\omega)^{4}}+\mathcal{O}(e^{2}),\end{equation}
which evidently proves that $B/R^{2}=\mathcal{O}(D/z)$ is a small
quantity justifying a posteriori our series expansions. In addition,
since~$B$ is relatively small but finite, the term~$z$ in the
action~(\ref{eq:fRQ-ansatz}) does not diverge.

Finally, let us determine the coefficient~$e_{a}$ in the scale factor
correction term by comparing~$R$ in (\ref{eq:Roft-early-full})
with Eq.~(\ref{eq:kR-early}), yielding~$e_{a}=-\frac{10}{63}$
for dust ($\omega=0$) and $e_{a}=-\frac{817}{6160}$ for radiation
($\omega=\frac{1}{3}$), respectively. Note that the term~$e\frac{9}{4}\frac{D}{mz}$
in $R$ does not come from the first-order corrections $\propto e_{1,2,a}$
in Eqs.~(\ref{eq:Om-early}) and~(\ref{eq:h00-early}), but it emerges
as the leading term of~$R_{i}$ and it provides the correct (non-zero)
value for~$B$. With these results one can check explicitely that
$z=\beta(R^{(2/3)}/B)^{m}$ and the validity of the EOMs in Eqs.~(\ref{eq:P-EOM-1})
and~(\ref{eq:P-EOM-2}).

\subsection{Late universe}

\label{sub:Cosmo-Late}Analogously to the discussion on the early
universe in the previous section, we determine now the scale factor
evolution at late times, when $r=\rho+p\ll|\rho_{e}|$ and~$\rho_{e}<0$.
After plugging~$B$ and~$B/R^{2}$ from Eqs.~(\ref{eq:B-late})
and~(\ref{eq:BR2-late}) into the expressions~(\ref{eq:Om-Cosmo})
and~(\ref{eq:h00-Cosmo}) for the auxiliary metric~$h_{ab}$, we
find\begin{eqnarray}
\Omega & = & c_{2}\rho_{e}^{-\frac{3}{2}}\left(1+e\, e_{1}\,\frac{1}{7}\frac{r}{\rho_{e}}+\mathcal{O}(e^{2})\right),\\
h_{00} & = & c_{2}\left(-\frac{4}{3}mz/e\right)\rho_{e}^{-\frac{5}{2}}\left(1-e\, e_{2}\,\frac{3}{7}\frac{r}{\rho_{e}}+\mathcal{O}(e^{2})\right).\end{eqnarray}
Also here, the minus sign in~$\rho_{e}$ should be absorbed in the
constant~$c_{2}=\sqrt{\pm\frac{27}{2}(\frac{2}{9}mz)^{3}(-\kappa)^{2}}$,
which drops out when calculating the connection~$\Gamma(h)$ in~(\ref{eq:LC-Con-h}).
Like before~$e=1$ represents powers of suppressed terms like $B/R^{2}=\mathcal{O}(\rho_{e}/z)$
or~$r/\rho_{e}$, and $e_{1,2}=1$ signal first-order corrections.
The non-vanishing components of the connection are given by (with
$i=1,2,3$)\begin{eqnarray}
\Gamma_{00}^{0} & = & -e\, e_{2}\,\frac{3}{14}\frac{\dot{r}}{\rho_{e}}+\mathcal{O}(e^{2}),\\
\Gamma_{ii}^{0} & = & e\,\frac{3a^{2}}{4mz}\rho_{e}\frac{\dot{a}}{a}+\mathcal{O}(e^{2}),\\
\Gamma_{i0}^{i} & = & \frac{\dot{a}}{a}+e\, e_{1}\,\frac{1}{14}\frac{\dot{r}}{\rho_{e}}+\mathcal{O}(e^{2}).\end{eqnarray}
Furthermore, we assume a constant EOS~$\omega$ for the matter content
as in Eq.~(\ref{eq:rhoOfa}), which simplifies the expressions for
the Ricci tensor components~$R_{0}=R_{0}^{\,\,0}$ and $R_{i}=R_{1}^{\,\,1}$
from Eq.~(\ref{eq:P-RicciTensor}),\begin{eqnarray}
R_{0} & = & 3\frac{\ddot{a}}{a}+e\,\frac{9}{14}(1+\omega)\frac{r}{\rho_{e}}\left(\left(e_{1}(2+3\omega)-3e_{2}\right)\left(\frac{\dot{a}}{a}\right)^{2}-e_{1}\frac{\ddot{a}}{a}\right)+\mathcal{O}(e^{2}),\nonumber \\
R_{i} & = & e\frac{3\rho_{e}}{4mz}\left(2\left(\frac{\dot{a}}{a}\right)^{2}+\frac{\ddot{a}}{a}\right)+\mathcal{O}(e^{2}).\end{eqnarray}
As before, the scale factor behaviour can be found by comparing the
zero-order term in the Ricci scalar~$R=R_{0}+3R_{i}=3\frac{\ddot{a}}{a}+\mathcal{O}(e)$
with Eq.~(\ref{eq:kR-late}) in the limit~$r\rightarrow0$, \begin{equation}
\kappa\,3\frac{\ddot{a}}{a}=\rho_{e}.\end{equation}
The solution is a linear combination of exponential functions, $a(t)\propto\exp(\pm H_{e}t)$
with constant~$H_{e}$. Since the shrinking solution quickly decays
at late times we drop it and consider only the de~Sitter-like solution\begin{equation}
a(t)\propto\exp(H_{e}t)\left(1+e\, e_{a}\frac{r}{\rho_{e}}\right),\label{eq:Late-aoft}\end{equation}
where a first-order correction proportional to the constant~$e_{a}$
has been added. Consequently,\[
H=H_{e}\left(1+e\, e_{a}\frac{-3(1+\omega)r}{\rho_{e}}+\mathcal{O}(e^{2})\right),\,\,\,\frac{\ddot{a}}{a}=H_{e}^{2}\left(1+e\, e_{a}\frac{3(1+3\omega)(1+\omega)r}{\rho_{e}}+\mathcal{O}(e^{2})\right),\]
\begin{equation}
R_{0}=3H_{e}^{2}\left(1+e\,\frac{3(3\omega-2+14e_{a}(1+3\omega))(1+\omega)r}{14\rho_{e}}+\mathcal{O}(e^{2})\right),\,\,\, R_{i}=e\,\frac{9H_{e}^{2}\rho_{e}}{4mz}+\mathcal{O}(e^{2}),\end{equation}
where~$e_{1,2}=1$ was applied. Neglecting the tiny $\mathcal{O}(\rho_{e}/z)$-term
in~$R_{i}$, the Ricci scalar~$R=R_{0}+3R_{i}$ will be consistent
with Eq.~(\ref{eq:kR-late}), $\kappa R=\rho_{e}-\frac{3}{7}r$,
if\begin{equation}
\rho_{e}=3\kappa H_{e}^{2}\,\,\,\text{and}\,\,\, e_{a}=\frac{-\omega}{14(1+\omega)}.\label{eq:Late-He-rhoe}\end{equation}
From the first condition the parameter~$\beta$ in $L_{3}=\rho_{e}^{7}<0$
can be determined because~$-\kappa\sim1/G_{N}$ was suggested in
Sec.~\ref{sub:Cosmo-Early} and thus $-\rho_{e}\sim-\kappa H_{e}^{2}$
must be of the order of the late-time critical energy density. This
is just the effective CC corresponding to the de~Sitter solution
(\ref{eq:Late-aoft}) with the tiny observed Hubble rate~$H_{e}$.
For dust matter~($\omega=0$) the scale factor correction~$\propto e_{a}$
vanishes and the expansion is purely de~Sitter-like.

As in the previous section, the squared Ricci tensor reads~$Q=(R_{0})^{2}+\mathcal{O}(e^{2})$
and we find that the denominator of~$z$ in Eq.~(\ref{eq:fRQ-ansatz})
is highly suppressed but non-zero,\begin{equation}
B=R^{2}-Q=6R\cdot R_{i}-12(R_{i})^{2}=e\frac{81H_{e}^{4}\rho_{e}}{2mz}+\mathcal{O}(e^{2}).\label{eq:Late-Boft}\end{equation}
Therefore, our expansion in $B/R^{2}=\mathcal{O}(\rho_{e}/z)$ turns
out to be justified also at late times.

To finalise this section we estimate the magnitude of the parameter~$\beta$
expressed as a power of an energy scale~$M_{\beta}$. The dimensional
analysis of~$z$ in Eq.~(\ref{eq:fRQ-ansatz}) implies $|\beta|=M_{\beta}^{d}$
with the exponent given by $d=4+\frac{8}{3}m$. From Eq.~(\ref{eq:z})
we know that~$z$ is of the order of the large cosmological term~$\Lambda$,
and in Eq.~(\ref{eq:Late-He-rhoe}) we related $(-\rho_{e})\sim(10^{-12}\,\text{GeV})^{4}$
to the tiny observed energy density of the effective late-time CC.
Finally, in Sec.~\ref{sub:Cosmo-Early} the parameter $(-\kappa)\sim(10^{18}\,\text{GeV})^{2}$
was fixed by the inverse Newton constant, and now we are prepared
to estimate~$\beta$ for given values of~$\Lambda$ and~$m$ by
using~$\rho_{e}^{7}=L_{3}$ in Eq.~(\ref{eq:kRpolyEqu}),\begin{equation}
\beta=\left(\frac{\rho_{e}^{7}}{\kappa^{4}}\right)^{\frac{m}{3}}\left(\frac{9}{2m}\right)^{m}z^{(1-m)}.\label{eq:beta-parameter}\end{equation}
First, we note that with large values of $m$ the tiny first factor
in~$\beta$ produces small values of~$M_{\beta}=|\beta|^{1/d}$,
which makes this parameter range less attractive. On the other hand,~$m<1$
provides an interesting range of energy scales. For instance, for
the vacuum energy density $|\Lambda|\sim|z|\sim(10^{16}\,\text{GeV})^{4}$
of a typical grand unified theory we obtain the following magnitudes
of $M_{\beta}$ in units of $\text{GeV}$: $10^{-51}$ for $m=3$,
$10^{-2}$ ($m=\frac{1}{3}$), $10^{4}$ ($m=\frac{1}{5}$), $10^{12}$
($m=\frac{1}{17}$), and $M_{\beta}$ approaches $z$ in the limit
$m\rightarrow0$. Apart from that, the structure of Eq.~(\ref{eq:beta-parameter})
does not require~$\beta$ to be fixed very precisely, which constitutes
a much better situation compared to the counterterm method in Eq.~(\ref{eq:counterterm}).

\section{Kottler solution}

\label{sec:Kottler}In the previous section we have seen that the
universe approaches a de~Sitter cosmos in the limit of vanishing
matter, where~$r\rightarrow0$ and~$\kappa R\rightarrow\rho_{e}$
in Eq.~(\ref{eq:kR-late}). We use this result now for discussing
the Kottler (Schwarzschild-de~Sitter) solution, which describes a
Schwarzschild black hole in the presence of a positive CC. For analysing
the CC filter in this environment we need the $4$-velocity vector~$u_{m}$
from Eq.~(\ref{eq:h_mn}), which we know already in the cosmological
setup. Therefore, it is useful to apply a transformation from the
spatially flat FLRW coordinates~$(t,\rho,\theta,\phi)$ with line
element \begin{equation}
ds^{2}=-dt^{2}+a^{2}(t)\left(d\rho^{2}+\rho^{2}d\theta^{2}+\rho^{2}\sin^{2}\theta d\phi^{2}\right)\label{eq:Kott-FRW-metric}\end{equation}
into Lema\^itre coordinates~$(\tau,\sigma,\theta,\phi)$, where
the angles~$\theta$ and~$\phi$ remain untouched,\begin{equation}
ds^{2}=-A\, d\tau^{2}+A^{-1}d\sigma^{2}+\sigma^{2}d\theta^{2}+\sigma^{2}\sin^{2}\theta d\phi^{2}.\label{eq:Kott-Lemaitre-metric}\end{equation}
In the latter metric the mass of the black hole in terms of the Schwarzschild
radius~$r_{s}$ and respectively the de~Sitter radius~$r_{e}$
define the term\begin{equation}
A(\sigma)=1-\frac{r_{s}}{\sigma}-\frac{\sigma^{2}}{r_{e}^{2}}.\label{eq:Kott-A}\end{equation}
The radial coordinates are related by~$\sigma=a(t)\rho$ and if we
require that~$A=1-(H\sigma)^{2}$ with the Hubble rate~$H=\frac{\dot{a}}{a}$,
the transformation rules between both metrics in~(\ref{eq:Kott-FRW-metric})
and~(\ref{eq:Kott-Lemaitre-metric}) read\begin{equation}
\frac{\partial\sigma}{\partial t}=H\sigma=\sqrt{1-A},\,\frac{\partial\sigma}{\partial\rho}=a,\,\frac{\partial\tau}{\partial t}=\frac{1}{A},\,\frac{\partial\tau}{\partial\rho}=\frac{aH\sigma}{A}=\frac{a\sqrt{1-A}}{A}.\label{eq:Kott-Trafo}\end{equation}
Next, we consider the $4$-velocity vector~$u_{m}$ in~(\ref{eq:h_mn}),
which has only one non-vanishing component~$u_{t}=1$ in FLRW coordinates.
Via the relations~(\ref{eq:Kott-Trafo}) the corresponding components
in Lema\^itre coordinates can be obtained easily,\begin{equation}
u_{\tau}=1,\, u_{\sigma}=-\frac{H\sigma}{A}=-\frac{\sqrt{1-A}}{A}.\label{eq:Kott-u_tau-r}\end{equation}
Of course, the norm $u_{m}u^{m}=-1$ remains invariant under this
change. With Eqs.~(\ref{eq:Kott-Lemaitre-metric}) and~(\ref{eq:Kott-u_tau-r})
the auxiliary metric in~(\ref{eq:h_mn}) is given by\begin{eqnarray}
h_{\tau\tau} & = & \Omega\left(g_{\tau\tau}-L_{u}u_{\tau}u_{\tau}\right)=\Omega\,(-A-L_{u})\label{eq:Kott-h_mn}\\
h_{\sigma\sigma} & = & \Omega\left(g_{\sigma\sigma}-L_{u}u_{\sigma}u_{\sigma}\right)=\Omega\left(A^{-1}-L_{u}\frac{1-A}{A^{2}}\right),\nonumber \\
h_{\tau\sigma} & = & \Omega\left(g_{\tau\sigma}-L_{u}u_{\tau}u_{\sigma}\right)=\Omega\left(0+L_{u}\frac{\sqrt{1-A}}{A}\right),\nonumber \\
h_{mn} & = & \Omega\, g_{mn}\,\,\,\text{for the other components.}\nonumber \end{eqnarray}
Interestingly, $h_{mn}$ has non-zero off-diagonal elements, whereas
the physical metric~$g_{mn}$ does not. Here, we have introduced
the variable~$L_{u}$, which follows from Eqs.~(\ref{eq:L2-Sol}),~(\ref{eq:L1-L2-Sol})
and~(\ref{eq:BR2-late}), and its leading term reads\begin{equation}
L_{u}:=\frac{L_{2}}{L_{1}-L_{2}}=6\frac{R^{2}}{B}(1+\varepsilon)=6\,\frac{\frac{2}{9}mz}{\rho_{e}}(1+\varepsilon).\label{eq:Kott-Lu}\end{equation}
Since we consider the limit~$r\rightarrow0$, both terms~$\Omega$
and~$L_{u}$ are constant. Thus~$\Omega$ will drop out from the
connection~$\Gamma_{bc}^{a}$ according to Eq.~(\ref{eq:LC-Con-h}).
For the non-vanishing components we find\begin{eqnarray}
\Gamma_{\tau\tau}^{\tau} & = & -\frac{L_{u}}{1+L_{u}}\cdot\frac{\sqrt{1-A}}{2\sigma A}\left(\frac{r_{s}}{\sigma}+2\frac{\sigma^{2}}{r_{e}^{2}}\right)=A^{2}(1-A)\Gamma_{\sigma\sigma}^{\tau}=-\Gamma_{\tau\sigma}^{\sigma},\label{eq:Kott-Gammas}\\
\Gamma_{\tau\sigma}^{\tau} & = & -\frac{A+L_{u}(A-1)}{(1+L_{u})2\sigma A}\left(\frac{r_{s}}{\sigma}+2\frac{\sigma^{2}}{r_{e}^{2}}\right),\,\,\,\Gamma_{\theta\theta}^{\tau}=-\frac{L_{u}\sigma\sqrt{1-A}}{(1+L_{u})A}=\sin^{-2}\theta\,\Gamma_{\phi\phi}^{\tau}\nonumber \\
\Gamma_{\tau\tau}^{\sigma} & = & -\frac{A+L_{u}}{(1+L_{u})2\sigma}\left(\frac{r_{s}}{\sigma}+2\frac{\sigma^{2}}{r_{e}^{2}}\right),\,\,\,\Gamma_{\sigma\sigma}^{\sigma}=-\Gamma_{\tau\sigma}^{\tau},\,\,\,\Gamma_{\theta\theta}^{\sigma}=-\frac{\sigma(A+L_{u})}{1+L_{u}}=\sin^{-2}\theta\,\Gamma_{\phi\phi}^{\sigma}\nonumber \\
\Gamma_{\sigma\theta}^{\theta} & = & \frac{1}{\sigma}=\Gamma_{\sigma\phi}^{\phi},\,\,\,\Gamma_{\phi\phi}^{\theta}=-\cos\theta\sin\theta,\,\,\,\Gamma_{\phi\theta}^{\phi}=\cot\theta.\nonumber \end{eqnarray}
Moreover, the non-zero components of the Ricci tensor read\begin{eqnarray}
R_{\tau\tau} & = & -\frac{3(A+L_{u})}{r_{e}^{2}(1+L_{u})},\,\,\, R_{\tau\sigma}=R_{\sigma\tau}=\frac{3L_{u}\sqrt{1-A}}{r_{e}^{2}(1+L_{u})A},\label{eq:Kott-R_mn}\\
R_{\sigma\sigma} & = & -\frac{3(L_{u}(1-A)-A)}{r_{e}^{2}(1+L_{u})A^{2}},\,\,\, R_{\theta\theta}=\frac{3\sigma^{2}}{r_{e}^{2}(1+L_{u})}=\sin^{-2}\theta\, R_{\phi\phi}.\nonumber \end{eqnarray}
Also here we find non-diagonal entries, however, they belong to the
Palatini Ricci tensor, which is derived from~$h_{mn}$ and not from
the physical metric. Finally, we show the results for the scalar invariants\begin{eqnarray}
R & = & R_{ab}g^{ab}=\frac{3}{r_{e}^{2}}\cdot\frac{4+L_{u}}{1+L_{u}},\label{eq:Kott-R}\\
Q & = & R_{ab}R_{cd}g^{ac}g^{bd}=\frac{(R_{\tau\tau})^{2}}{A^{2}}+2(R_{01})^{2}\frac{A}{-A}+A^{2}(R_{\sigma\sigma})^{2}+\frac{(R_{\theta\theta})^{2}}{\sigma^{4}}+\frac{(R_{\phi\phi})^{2}}{\sin^{4}\theta\sigma^{4}}\nonumber \\
 & = & \left(\frac{3}{r_{e}^{2}}\right)^{2}\cdot\frac{4+2L_{u}+L_{u}^{2}}{(1+L_{u})^{2}},\label{eq:Kott-Q}\\
B & = & R^{2}-Q=\left(\frac{3}{r_{e}^{2}}\right)^{2}\cdot\frac{6(2+L_{u})}{(1+L_{u})^{2}}=R^{2}\frac{6}{L_{u}}(1+\mathcal{O}(\varepsilon)).\label{eq:Kott-B}\end{eqnarray}
At leading order the last equation is consistent with Eq.~(\ref{eq:BR2-late}),
and it implies that~$B/R^{2}=\mathcal{O}(\rho_{e}/z)\ll1$ is a suitable
expansion parameter. Moreover, by comparing~$R$ in Eq.~(\ref{eq:Kott-R})
with the late-time results from Eqs.~(\ref{eq:kR-late}) and~(\ref{eq:Late-He-rhoe}),
we find at leading order\begin{equation}
R=\frac{3}{r_{e}^{2}}=\frac{\rho_{e}}{\kappa}=3H_{e}^{2}.\label{eq:Kott-r0-He}\end{equation}
This implies that the de~Sitter radius~$r_{e}$ as a parameter in
the Kottler metric~(\ref{eq:Kott-Lemaitre-metric}) is given by the
inverse of the final Hubble rate~$H_{e}=r_{e}^{-1}$, just as in
general relativity. Remember that~$H_{e}$ originates from the small
effective vacuum energy density of the order~$|\rho_{e}|$ and not
from the large CC~$\Lambda$. Note also that Eqs.~(\ref{eq:Kott-u_tau-r})
and (\ref{eq:Kott-r0-He}) are sufficient to show that the metric~(\ref{eq:Kott-Lemaitre-metric})
with $A$ given in~(\ref{eq:Kott-A}) is a solution of our Palatini
model. The coordinate transformation~(\ref{eq:Kott-Trafo}) just
served us to obtain the vector~$u_{m}$ in Lema\^itre coordinates.
As a result of this section, the CC is relaxed also in situations,
which can be well described by the Kottler metric. This can be very
useful for solar system tests and the construction of vacuole solutions
of the Einstein-Strau\ss{} type~\cite{Einstein:1945id,Balbinot:1988zc},
which we would like to discuss in the future. For the vacuole solutions
we cannot assume $r\ll|\rho_{e}|$ as we did in Sec.~\ref{sub:Cosmo-Late},
but the matter density~$r$ must be treated on equal footing with
the effective vacuum energy density~$|\rho_{e}|$.

\section{Conclusions and outlook}

\label{sec:Conclusions}In the context of the old CC problem, we have
presented a modified gravity model, which filters out a large CC~$\Lambda$
independent of its origin. Thanks to the Palatini formalism, we avoid
problems coming from extra degrees of freedom, which are often present
in the metric formalism. In this work, several aspects of our filter
scenario have been analysed with the result that the standard Big
Bang history of the universe is not prevented by a large CC term even
when it dominates in size over matter or radiation.

Unlike many models for cosmological late-time acceleration, our setup
in Eq.~(\ref{eq:fRQ-ansatz}) does not represent a small correction
to the Einstein-Hilbert term because the term~$z$ containing the
Ricci scalar~$R$ and the squared Ricci tensor~$Q$, plays a crucial
role for the filter effect. Thus, it is neither useful nor necessary
to consider the limit~$z\rightarrow0$. Furthermore, we have shown
that the Hubble expansion rate in the early universe is dominated
by the (non-vacuum) matter energy density, whereas CC contributions
with equation of state~$\omega=-1$ do not contribute at all in leading
order. This filter effect implies a cosmological background evolution
similar to general relativity, where the large CC has been removed
somehow. We have found reasonable results in the matter and radiation
eras. In the latter the effective Newton constant is slightly increased
but consistent with recent bounds~\cite{Steigman:2010zz}. At late
times, when matter has diluted away, we enter a de~Sitter phase,
where the expansion rate~$H_{e}$ is not dominated by the large CC,
but instead it depends only on the magnitudes of the parameter~$\beta$
and~$\Lambda$ in Eq.~(\ref{eq:kRpolyEqu}). Hence, the inclusion
of a CC counterterm which cancels~$\Lambda$ with extremely high
precision is not necessary for describing the currently observed accelerated
expansion. Our cosmological results are supported by the existence
of a black hole solution of the Kottler type, in which the effective
CC parameter complies with the value of the final de~Sitter expansion
rate~$H_{e}$. This suggests that cosmology and the astro-physical
domain can be smoothly connected.

The robustness of the CC filter can be seen also from a different
perspective. So far we have considered~$\Lambda$ to be a large constant
generally of the order of the largest contribution to the CC. During
the cosmic evolution, however, it is not unlikely that shifts~$\delta\Lambda$
of vacuum energy occur, as a result of phase transitions for instance.
We have shown that these contributions, which are smaller than~$\Lambda$
by definition, have no influence on gravity at leading order if we
treat them as part of the matter sector described by~$r=\rho+p$
in Eq.~(\ref{eq:SeqZeroRelax}). Alternatively, one could shift the
cosmological term by $\Lambda\rightarrow\Lambda+\delta\Lambda$, but
also this procedure does not change our results because the differences
are of the order~$|\delta\Lambda/\Lambda|\ll1$ contributing only
to the small correction terms. Hence, the exact value of~$\Lambda$
is not important for the CC filter effect even when vacuum shifts~$\delta\Lambda$
arise dynamically as the universe evolves. Note that~$\delta\Lambda$
can be much larger than the current critical energy density despite
being small compared to~$\Lambda$.

Let us now sketch some open points and discuss possible generalisations
of our setup, which we want to address in the future. First, it is
important to mention that our filter effect is based on the largeness
of the CC and its universal energy-momentum structure. This can be
seen nicely in Eq.~(\ref{eq:S3}), where the CC contribution to the
Ricci scalar~$R$ can be completely removed by a suitable model building
construction, which leads to Eq.~(\ref{eq:SeqZeroRelax}). It is
clear that as long as~$\Lambda$ dominates over other sources in
the energy-momentum tensor and its trace~(\ref{eq:T-stress}), the
procedure will be the same, which removes from~$R$ the dominating
term~$z\sim\Lambda$ as given in~Eq.~(\ref{eq:z}). Therefore,
we can at least conjecture that matter sources different from the
perfect fluid form as well as other background metrics exhibit also
the CC filter property. This would be useful for future studies of
the Newtonian limit as well as perturbations in cosmology and astro-physical
setups, which are more involved than the Kottler solution. For the
latter, it seems that vacuole space-times can be easily constructed
once a solution is found which interpolates between the matter era
and the final de~Sitter phase. The smoothness of~$R$ as shown in
Fig.~\ref{fig:kRofr-m3} suggests such a solution. More difficult
could be the analysis of the very early cosmological epoch when matter
dominated over~$\Lambda$ (if there was such a time) or during primordial
inflation, respectively. In that case many approximations we used
for the subsequent eras cannot be applied anymore, and a new analysis
is necessary.

Apart from that, we have shown only one numerical example to support
our analytical considerations, and it would be nice to have more numerical
results discussing e.g.\ the transition phases between different
epochs and the constraints on the parameters~$\beta$ and~$m$.
Also the~$L_{3}>0$ case in Eq.~(\ref{eq:kRpolyEqu}) might correspond
to an interesting solution, maybe it describes a contracting universe
at late times because a vanishing matter density is not a solution
of Eq.~(\ref{eq:kRpolyEqu}). However, since we observe an accelerating
cosmos, the~$L_{3}<0$ case discussed here is preferred.

Moreover, we should mention that Palatini models are subject to constraints
from astrophysical bodies and even small scale atomic physics, see
e.g.~\cite{Barausse:2007pn,Barausse:2007ys,Li:2008bma,Li:2008fa,Sotiriou:2008rp}.
However, many results correspond to $f(R)$-type actions or to models,
where the Einstein-Hilbert term is amended by a correction, which
becomes large for low curvature. It is clear that our setup is not
of this type. We will see in the future whether this is an advantage
or not. Nevertheless, it might help that the Ricci scalar in Eq.~(\ref{eq:kR-late})
and the denominator function~\textbf{$B$ }in~(\ref{eq:B-late})
approach a non-zero constant in the limit of vanishing matter. This
suggests a stable vacuum, which limits the values of the geometrical
scalars from below. Finally, it has been argued in Refs.~\cite{Li:2008bma,Li:2008fa}
that the averaging over matter sources in the Palatini framework has
some non-trivial aspects, too.

In the end, it would be interesting to know whether the CC filter
effect is just a curiosity of our specific model in Eq.~(\ref{eq:fRQ-ansatz}),
or if there are completely different choices, which have the same
property. We want address these questions in the future.

\subsection*{Acknowledgements}

I would like to thank Joan Sol\`a and Hrvoje \v{S}tefan\v{c}i\'{c}
for useful discussions and comments.\\
This work has been supported in part by MEC and FEDER under project
FPA2007-66665, by the Spanish Consolider-Ingenio 2010 program CPAN
CSD2007-00042 and by DIUE/CUR Generalitat de Catalunya under project
2009SGR502.

{\small \bibliographystyle{utphys}
\bibliography{Palatini-Relax}

\providecommand{\href}[2]{#2}\begingroup\raggedright\begin{thebibliography}{10}

\bibitem{Bauer:2010wj}
F.~Bauer, J.~Sola, and H.~Stefancic, ``{Dynamically avoiding fine-tuning the
  cosmological constant: the 'Relaxed Universe'},''
  \href{http://dx.doi.org/10.1088/1475-7516/2010/12/029}{{\em JCAP} {\bfseries
  1012} (2010) 029},
\href{http://arxiv.org/abs/1006.3944}{{\ttfamily arXiv:1006.3944 [hep-th]}}.
%%CITATION = 1006.3944;%%.

\bibitem{Weinberg:1988cp}
S.~Weinberg, ``{The cosmological constant problem},''
\href{http://dx.doi.org/10.1103/RevModPhys.61.1}{{\em Rev. Mod. Phys.}
  {\bfseries 61} (1989) 1--23}.
%%CITATION = RMPHA,61,1;%%.

\bibitem{Bertolami:2009nr}
O.~Bertolami, ``{The cosmological constant problem: a user's guide},''
  \href{http://dx.doi.org/10.1142/S0218271809015862}{{\em Int. J. Mod. Phys.}
  {\bfseries D18} (2009) 2303--2310},
\href{http://arxiv.org/abs/0905.3110}{{\ttfamily arXiv:0905.3110 [gr-qc]}}.
%%CITATION = 0905.3110;%%.

\bibitem{Spergel:2006hy}
{\bfseries WMAP} Collaboration, D.~N. Spergel {\em et al.}, ``{Wilkinson
  Microwave Anisotropy Probe (WMAP) three year results: Implications for
  cosmology},'' \href{http://dx.doi.org/10.1086/513700}{{\em Astrophys. J.
  Suppl.} {\bfseries 170} (2007) 377},
\href{http://arxiv.org/abs/astro-ph/0603449}{{\ttfamily
  arXiv:astro-ph/0603449}}.
%%CITATION = ASTRO-PH/0603449;%%.

\bibitem{Knop:2003iy}
{\bfseries Supernova Cosmology Project} Collaboration, R.~A. Knop {\em et al.},
  ``{New Constraints on $\Omega_M$, $\Omega_\Lambda$, and w from an Independent
  Set of Eleven High-Redshift Supernovae Observed with HST},''
  \href{http://dx.doi.org/10.1086/378560}{{\em Astrophys. J.} {\bfseries 598}
  (2003) 102},
\href{http://arxiv.org/abs/astro-ph/0309368}{{\ttfamily
  arXiv:astro-ph/0309368}}.
%%CITATION = ASTRO-PH/0309368;%%.

\bibitem{Riess:2004nr}
{\bfseries Supernova Search Team} Collaboration, A.~G. Riess {\em et al.},
  ``{Type Ia Supernova Discoveries at z>1 From the Hubble Space Telescope:
  Evidence for Past Deceleration and Constraints on Dark Energy Evolution},''
  \href{http://dx.doi.org/10.1086/383612}{{\em Astrophys. J.} {\bfseries 607}
  (2004) 665--687},
\href{http://arxiv.org/abs/astro-ph/0402512}{{\ttfamily
  arXiv:astro-ph/0402512}}.
%%CITATION = ASTRO-PH/0402512;%%.

\bibitem{Carroll:2000fy}
S.~M. Carroll, ``{The cosmological constant},'' {\em Living Rev. Rel.}
  {\bfseries 4} (2001) 1,
\href{http://arxiv.org/abs/astro-ph/0004075}{{\ttfamily
  arXiv:astro-ph/0004075}}.
%%CITATION = ASTRO-PH/0004075;%%.

\bibitem{Peebles:2002gy}
P.~J.~E. Peebles and B.~Ratra, ``{The cosmological constant and dark energy},''
  \href{http://dx.doi.org/10.1103/RevModPhys.75.559}{{\em Rev. Mod. Phys.}
  {\bfseries 75} (2003) 559--606},
\href{http://arxiv.org/abs/astro-ph/0207347}{{\ttfamily
  arXiv:astro-ph/0207347}}.
%%CITATION = ASTRO-PH/0207347;%%.

\bibitem{Padmanabhan:2002ji}
T.~Padmanabhan, ``{Cosmological constant: The weight of the vacuum},''
  \href{http://dx.doi.org/10.1016/S0370-1573(03)00120-0}{{\em Phys. Rept.}
  {\bfseries 380} (2003) 235--320},
\href{http://arxiv.org/abs/hep-th/0212290}{{\ttfamily arXiv:hep-th/0212290}}.
%%CITATION = HEP-TH/0212290;%%.

\bibitem{Nojiri:2006ri}
S.~Nojiri and S.~D. Odintsov, ``{Introduction to modified gravity and
  gravitational alternative for dark energy},''
  \href{http://dx.doi.org/10.1142/S0219887807001928}{{\em ECONF} {\bfseries
  C0602061} (2006) 06},
\href{http://arxiv.org/abs/hep-th/0601213}{{\ttfamily arXiv:hep-th/0601213}}.
%%CITATION = HEP-TH/0601213;%%.

\bibitem{Copeland:2006wr}
E.~J. Copeland, M.~Sami, and S.~Tsujikawa, ``{Dynamics of dark energy},''
  \href{http://dx.doi.org/10.1142/S021827180600942X}{{\em Int. J. Mod. Phys.}
  {\bfseries D15} (2006) 1753--1936},
\href{http://arxiv.org/abs/hep-th/0603057}{{\ttfamily arXiv:hep-th/0603057}}.
%%CITATION = HEP-TH/0603057;%%.

\bibitem{Stefancic:2008zz}
H.~Stefancic, ``{The solution of the cosmological constant problem from the
  inhomogeneous equation of state - a hint from modified gravity?},''
  \href{http://dx.doi.org/10.1016/j.physletb.2008.10.065}{{\em Phys. Lett.}
  {\bfseries B670} (2009) 246--253},
\href{http://arxiv.org/abs/0807.3692}{{\ttfamily arXiv:0807.3692 [gr-qc]}}.
%%CITATION = 0807.3692;%%.

\bibitem{Grande:2006nn}
J.~Grande, J.~Sola, and H.~Stefancic, ``{LXCDM: a cosmon model solution to the
  cosmological coincidence problem?},''
  \href{http://dx.doi.org/10.1088/1475-7516/2006/08/011}{{\em JCAP} {\bfseries
  0608} (2006) 011},
\href{http://arxiv.org/abs/gr-qc/0604057}{{\ttfamily arXiv:gr-qc/0604057}}.
%%CITATION = GR-QC/0604057;%%.

\bibitem{Bauer:2009ke}
F.~Bauer, J.~Sola, and H.~Stefancic, ``{Relaxing a large cosmological
  constant},'' \href{http://dx.doi.org/10.1016/j.physletb.2009.06.065}{{\em
  Phys. Lett.} {\bfseries B678} (2009) 427--433},
\href{http://arxiv.org/abs/0902.2215}{{\ttfamily arXiv:0902.2215 [hep-th]}}.
%%CITATION = 0902.2215;%%.

\bibitem{Bauer:2009jk}
F.~Bauer, ``{Perturbations in the relaxation mechanism for a large cosmological
  constant},'' \href{http://dx.doi.org/10.1088/0264-9381/27/5/055001}{{\em
  Class. Quant. Grav.} {\bfseries 27} (2010) 055001},
\href{http://arxiv.org/abs/0909.2237}{{\ttfamily arXiv:0909.2237 [gr-qc]}}.
%%CITATION = 0909.2237;%%.

\bibitem{Bonanno:2001hi}
A.~Bonanno and M.~Reuter, ``{Cosmology with self-adjusting vacuum energy
  density from a renormalization group fixed point},''
  \href{http://dx.doi.org/10.1016/S0370-2693(01)01522-2}{{\em Phys. Lett.}
  {\bfseries B527} (2002) 9--17},
\href{http://arxiv.org/abs/astro-ph/0106468}{{\ttfamily
  arXiv:astro-ph/0106468}}.
%%CITATION = ASTRO-PH/0106468;%%.

\bibitem{Nobbenhuis:2004wn}
S.~Nobbenhuis, ``{Categorizing Different Approaches to the Cosmological
  Constant Problem},'' \href{http://dx.doi.org/10.1007/s10701-005-9042-8}{{\em
  Found. Phys.} {\bfseries 36} (2006) 613--680},
\href{http://arxiv.org/abs/gr-qc/0411093}{{\ttfamily arXiv:gr-qc/0411093}}.
%%CITATION = GR-QC/0411093;%%.

\bibitem{Barr:2006mp}
S.~M. Barr, S.-P. Ng, and R.~J. Scherrer, ``{Classical cancellation of the
  cosmological constant re- considered},''
  \href{http://dx.doi.org/10.1103/PhysRevD.73.063530}{{\em Phys. Rev.}
  {\bfseries D73} (2006) 063530},
\href{http://arxiv.org/abs/hep-ph/0601053}{{\ttfamily arXiv:hep-ph/0601053}}.
%%CITATION = HEP-PH/0601053;%%.

\bibitem{Diakonos:2007au}
F.~K. Diakonos and E.~N. Saridakis, ``{A Statistical Solution to the
  Cosmological Constant Problem in the Brane world},''
  \href{http://dx.doi.org/10.1088/1475-7516/2009/02/030}{{\em JCAP} {\bfseries
  0902} (2009) 030},
\href{http://arxiv.org/abs/0708.3143}{{\ttfamily arXiv:0708.3143 [hep-th]}}.
%%CITATION = 0708.3143;%%.

\bibitem{Klinkhamer:2007pe}
F.~R. Klinkhamer and G.~E. Volovik, ``{Self-tuning vacuum variable and
  cosmological constant},''
  \href{http://dx.doi.org/10.1103/PhysRevD.77.085015}{{\em Phys. Rev.}
  {\bfseries D77} (2008) 085015},
\href{http://arxiv.org/abs/0711.3170}{{\ttfamily arXiv:0711.3170 [gr-qc]}}.
%%CITATION = 0711.3170;%%.

\bibitem{Batra:2008cc}
P.~Batra, K.~Hinterbichler, L.~Hui, and D.~N. Kabat, ``{Pseudo-redundant vacuum
  energy},'' \href{http://dx.doi.org/10.1103/PhysRevD.78.043507}{{\em Phys.
  Rev.} {\bfseries D78} (2008) 043507},
\href{http://arxiv.org/abs/0801.4526}{{\ttfamily arXiv:0801.4526 [hep-th]}}.
%%CITATION = 0801.4526;%%.

\bibitem{Dvali:2007kt}
G.~Dvali, S.~Hofmann, and J.~Khoury, ``{Degravitation of the cosmological
  constant and graviton width},''
  \href{http://dx.doi.org/10.1103/PhysRevD.76.084006}{{\em Phys. Rev.}
  {\bfseries D76} (2007) 084006},
\href{http://arxiv.org/abs/hep-th/0703027}{{\ttfamily arXiv:hep-th/0703027}}.
%%CITATION = HEP-TH/0703027;%%.

\bibitem{Patil:2008sp}
S.~P. Patil, ``{Degravitation, Inflation and the Cosmological Constant as an
  Afterglow},'' \href{http://dx.doi.org/10.1088/1475-7516/2009/01/017}{{\em
  JCAP} {\bfseries 0901} (2009) 017},
\href{http://arxiv.org/abs/0801.2151}{{\ttfamily arXiv:0801.2151 [hep-th]}}.
%%CITATION = 0801.2151;%%.

\bibitem{Demir:2009be}
D.~A. Demir, ``{Vacuum Energy as the Origin of the Gravitational Constant},''
  \href{http://dx.doi.org/10.1007/s10701-009-9364-z}{{\em Found. Phys.}
  {\bfseries 39} (2009) 1407--1425},
\href{http://arxiv.org/abs/0910.2730}{{\ttfamily arXiv:0910.2730 [hep-th]}}.
%%CITATION = 0910.2730;%%.

\bibitem{Hassan:2010ys}
S.~F. Hassan, S.~Hofmann, and M.~von Strauss, ``{Brane Induced Gravity, its
  Ghost and the Cosmological Constant Problem},''
\href{http://arxiv.org/abs/1007.1263}{{\ttfamily arXiv:1007.1263 [hep-th]}}.
%%CITATION = 1007.1263;%%.

\bibitem{Unruh:1988in}
W.~G. Unruh, ``{A unimodular theory of canonical quantum gravity},''
\href{http://dx.doi.org/10.1103/PhysRevD.40.1048}{{\em Phys. Rev.} {\bfseries
  D40} (1989) 1048}.
%%CITATION = PHRVA,D40,1048;%%.

\bibitem{Henneaux:1989zc}
M.~Henneaux and C.~Teitelboim, ``{The cosmological constant and general
  covariance},''
\href{http://dx.doi.org/10.1016/0370-2693(89)91251-3}{{\em Phys. Lett.}
  {\bfseries B222} (1989) 195--199}.
%%CITATION = PHLTA,B222,195;%%.

\bibitem{Ng:1990rw}
Y.~J. Ng and H.~van Dam, ``{Possible solution to the cosmological constant
  problem},''
\href{http://dx.doi.org/10.1103/PhysRevLett.65.1972}{{\em Phys. Rev. Lett.}
  {\bfseries 65} (1990) 1972--1974}.
%%CITATION = PRLTA,65,1972;%%.

\bibitem{Ng:1990xz}
Y.~J. Ng and H.~van Dam, ``{Unimodular theory of gravity and the cosmological
  constant},''
\href{http://dx.doi.org/10.1063/1.529283}{{\em J. Math. Phys.} {\bfseries 32}
  (1991) 1337--1340}.
%%CITATION = JMAPA,32,1337;%%.

\bibitem{Smolin:2009ti}
L.~Smolin, ``{The quantization of unimodular gravity and the cosmological
  constant problem},'' \href{http://dx.doi.org/10.1103/PhysRevD.80.084003}{{\em
  Phys. Rev.} {\bfseries D80} (2009) 084003},
\href{http://arxiv.org/abs/0904.4841}{{\ttfamily arXiv:0904.4841 [hep-th]}}.
%%CITATION = 0904.4841;%%.

\bibitem{Bauer:2009ea}
F.~Bauer, J.~Sola, and H.~Stefancic, ``{The Relaxed Universe: towards solving
  the cosmological constant problem dynamically from an effective action
  functional of gravity},''
  \href{http://dx.doi.org/10.1016/j.physletb.2010.04.029}{{\em Phys. Lett.}
  {\bfseries B688} (2010) 269--272},
\href{http://arxiv.org/abs/0912.0677}{{\ttfamily arXiv:0912.0677 [hep-th]}}.
%%CITATION = 0912.0677;%%.

\bibitem{Vollick:2003ic}
D.~N. Vollick, ``{On the viability of the Palatini form of 1/R gravity},''
  \href{http://dx.doi.org/10.1088/0264-9381/21/15/N01}{{\em Class. Quant.
  Grav.} {\bfseries 21} (2004) 3813--3816},
\href{http://arxiv.org/abs/gr-qc/0312041}{{\ttfamily arXiv:gr-qc/0312041}}.
%%CITATION = GR-QC/0312041;%%.

\bibitem{Allemandi:2004wn}
G.~Allemandi, A.~Borowiec, and M.~Francaviglia, ``{Accelerated cosmological
  models in Ricci squared gravity},''
  \href{http://dx.doi.org/10.1103/PhysRevD.70.103503}{{\em Phys. Rev.}
  {\bfseries D70} (2004) 103503},
\href{http://arxiv.org/abs/hep-th/0407090}{{\ttfamily arXiv:hep-th/0407090}}.
%%CITATION = HEP-TH/0407090;%%.

\bibitem{Sotiriou:2005xe}
T.~P. Sotiriou, ``{The nearly Newtonian regime in Non-Linear Theories of
  Gravity},'' \href{http://dx.doi.org/10.1007/s10714-006-0328-8}{{\em Gen. Rel.
  Grav.} {\bfseries 38} (2006) 1407--1417},
\href{http://arxiv.org/abs/gr-qc/0507027}{{\ttfamily arXiv:gr-qc/0507027}}.
%%CITATION = GR-QC/0507027;%%.

\bibitem{Barausse:2007ys}
E.~Barausse, T.~P. Sotiriou, and J.~C. Miller, ``{Curvature singularities,
  tidal forces and the viability of Palatini f(R) gravity},''
  \href{http://dx.doi.org/10.1088/0264-9381/25/10/105008}{{\em Class. Quant.
  Grav.} {\bfseries 25} (2008) 105008},
\href{http://arxiv.org/abs/0712.1141}{{\ttfamily arXiv:0712.1141 [gr-qc]}}.
%%CITATION = 0712.1141;%%.

\bibitem{Li:2007xw}
B.~Li, J.~D. Barrow, and D.~F. Mota, ``{The Cosmology of Ricci-Tensor-Squared
  Gravity in the Palatini Variational Approach},''
  \href{http://dx.doi.org/10.1103/PhysRevD.76.104047}{{\em Phys. Rev.}
  {\bfseries D76} (2007) 104047},
\href{http://arxiv.org/abs/0707.2664}{{\ttfamily arXiv:0707.2664 [gr-qc]}}.
%%CITATION = 0707.2664;%%.

\bibitem{Sotiriou:2008rp}
T.~P. Sotiriou and V.~Faraoni, ``{f(R) Theories Of Gravity},''
  \href{http://dx.doi.org/10.1103/RevModPhys.82.451}{{\em Rev. Mod. Phys.}
  {\bfseries 82} (2010) 451--497},
\href{http://arxiv.org/abs/0805.1726}{{\ttfamily arXiv:0805.1726 [gr-qc]}}.
%%CITATION = 0805.1726;%%.

\bibitem{Exirifard:2007da}
Q.~Exirifard and M.~M. Sheikh-Jabbari, ``{Lovelock Gravity at the Crossroads of
  Palatini and Metric Formulations},''
  \href{http://dx.doi.org/10.1016/j.physletb.2008.02.012}{{\em Phys. Lett.}
  {\bfseries B661} (2008) 158--161},
\href{http://arxiv.org/abs/0705.1879}{{\ttfamily arXiv:0705.1879 [hep-th]}}.
%%CITATION = 0705.1879;%%.

\bibitem{Tsujikawa:2007tg}
S.~Tsujikawa, K.~Uddin, and R.~Tavakol, ``{Density perturbations in f(R)
  gravity theories in metric and Palatini formalisms},''
  \href{http://dx.doi.org/10.1103/PhysRevD.77.043007}{{\em Phys. Rev.}
  {\bfseries D77} (2008) 043007},
\href{http://arxiv.org/abs/0712.0082}{{\ttfamily arXiv:0712.0082 [astro-ph]}}.
%%CITATION = 0712.0082;%%.

\bibitem{Bauer:2008zj}
F.~Bauer and D.~A. Demir, ``{Inflation with Non-Minimal Coupling: Metric vs.
  Palatini Formulations},''
  \href{http://dx.doi.org/10.1016/j.physletb.2008.06.014}{{\em Phys. Lett.}
  {\bfseries B665} (2008) 222--226},
\href{http://arxiv.org/abs/0803.2664}{{\ttfamily arXiv:0803.2664 [hep-ph]}}.
%%CITATION = 0803.2664;%%.

\bibitem{Borunda:2008kf}
M.~Borunda, B.~Janssen, and M.~Bastero-Gil, ``{Palatini versus metric
  formulation in higher curvature gravity},''
  \href{http://dx.doi.org/10.1088/1475-7516/2008/11/008}{{\em JCAP} {\bfseries
  0811} (2008) 008},
\href{http://arxiv.org/abs/0804.4440}{{\ttfamily arXiv:0804.4440 [hep-th]}}.
%%CITATION = 0804.4440;%%.

\bibitem{Capozziello:2009nq}
S.~Capozziello, M.~De~Laurentis, and V.~Faraoni, ``{A bird's eye view of
  f(R)-gravity},''
\href{http://arxiv.org/abs/0909.4672}{{\ttfamily arXiv:0909.4672 [gr-qc]}}.
%%CITATION = 0909.4672;%%.

\bibitem{DeFelice:2010aj}
A.~De~Felice and S.~Tsujikawa, ``{f(R) theories},'' {\em Living Rev. Rel.}
  {\bfseries 13} (2010) 3,
\href{http://arxiv.org/abs/1002.4928}{{\ttfamily arXiv:1002.4928 [gr-qc]}}.
%%CITATION = 1002.4928;%%.

\bibitem{Goenner:2010tr}
H.~F.~M. Goenner, ``{Alternative to the Palatini method: A new variational
  principle},'' \href{http://dx.doi.org/10.1103/PhysRevD.81.124019}{{\em Phys.
  Rev.} {\bfseries D81} (2010) 124019},
\href{http://arxiv.org/abs/1003.5532}{{\ttfamily arXiv:1003.5532 [gr-qc]}}.
%%CITATION = 1003.5532;%%.

\bibitem{Meng:2003sx}
X.-H. Meng and P.~Wang, ``{Gravitational potential in Palatini formulation of
  modified gravity},''
  \href{http://dx.doi.org/10.1023/B:GERG.0000036052.81522.fe}{{\em Gen. Rel.
  Grav.} {\bfseries 36} (2004) 1947--1954},
\href{http://arxiv.org/abs/gr-qc/0311019}{{\ttfamily arXiv:gr-qc/0311019}}.
%%CITATION = GR-QC/0311019;%%.

\bibitem{Sotiriou:2005cd}
T.~P. Sotiriou, ``{Constraining $f(R)$ gravity in the Palatini formalism},''
  \href{http://dx.doi.org/10.1088/0264-9381/23/4/012}{{\em Class. Quant. Grav.}
  {\bfseries 23} (2006) 1253--1267},
\href{http://arxiv.org/abs/gr-qc/0512017}{{\ttfamily arXiv:gr-qc/0512017}}.
%%CITATION = GR-QC/0512017;%%.

\bibitem{Li:2008bma}
B.~Li, D.~F. Mota, and D.~J. Shaw, ``{Indistinguishable Macroscopic Behaviour
  of Palatini Gravities and General Relativity},''
  \href{http://dx.doi.org/10.1088/0264-9381/26/5/055018}{{\em Class. Quant.
  Grav.} {\bfseries 26} (2009) 055018},
\href{http://arxiv.org/abs/0801.0603}{{\ttfamily arXiv:0801.0603 [gr-qc]}}.
%%CITATION = 0801.0603;%%.

\bibitem{Li:2008fa}
B.~Li, D.~F. Mota, and D.~J. Shaw, ``{Microscopic and Macroscopic Behaviors of
  Palatini Modified Gravity Theories},''
  \href{http://dx.doi.org/10.1103/PhysRevD.78.064018}{{\em Phys. Rev.}
  {\bfseries D78} (2008) 064018},
\href{http://arxiv.org/abs/0805.3428}{{\ttfamily arXiv:0805.3428 [gr-qc]}}.
%%CITATION = 0805.3428;%%.

\bibitem{Olmo:2009xy}
G.~J. Olmo, H.~Sanchis-Alepuz, and S.~Tripathi, ``{Dynamical Aspects of
  Generalized Palatini Theories of Gravity},''
  \href{http://dx.doi.org/10.1103/PhysRevD.80.024013}{{\em Phys. Rev.}
  {\bfseries D80} (2009) 024013},
\href{http://arxiv.org/abs/0907.2787}{{\ttfamily arXiv:0907.2787 [gr-qc]}}.
%%CITATION = 0907.2787;%%.

\bibitem{Barragan:2010qb}
C.~Barragan and G.~J. Olmo, ``{Isotropic and Anisotropic Bouncing Cosmologies
  in Palatini Gravity},''
  \href{http://dx.doi.org/10.1103/PhysRevD.82.084015}{{\em Phys. Rev.}
  {\bfseries D82} (2010) 084015},
\href{http://arxiv.org/abs/1005.4136}{{\ttfamily arXiv:1005.4136 [gr-qc]}}.
%%CITATION = 1005.4136;%%.

\bibitem{Olmo:2008nf}
G.~J. Olmo and P.~Singh, ``{Effective Action for Loop Quantum Cosmology a la
  Palatini},'' \href{http://dx.doi.org/10.1088/1475-7516/2009/01/030}{{\em
  JCAP} {\bfseries 0901} (2009) 030},
\href{http://arxiv.org/abs/0806.2783}{{\ttfamily arXiv:0806.2783 [gr-qc]}}.
%%CITATION = 0806.2783;%%.

\bibitem{Vitagliano:2010pq}
V.~Vitagliano, T.~P. Sotiriou, and S.~Liberati, ``{The dynamics of generalized
  Palatini Theories of Gravity},''
  \href{http://dx.doi.org/10.1103/PhysRevD.82.084007}{{\em Phys. Rev.}
  {\bfseries D82} (2010) 084007},
\href{http://arxiv.org/abs/1007.3937}{{\ttfamily arXiv:1007.3937 [gr-qc]}}.
%%CITATION = 1007.3937;%%.

\bibitem{Woodard:2006nt}
R.~P. Woodard, ``{Avoiding dark energy with 1/R modifications of gravity},''
  \href{http://dx.doi.org/10.1007/978-3-540-71013-4_14}{{\em Lect. Notes Phys.}
  {\bfseries 720} (2007) 403--433},
\href{http://arxiv.org/abs/astro-ph/0601672}{{\ttfamily
  arXiv:astro-ph/0601672}}.
%%CITATION = ASTRO-PH/0601672;%%.

\bibitem{Bekenstein:1992pj}
J.~D. Bekenstein, ``{The Relation between physical and gravitational
  geometry},'' \href{http://dx.doi.org/10.1103/PhysRevD.48.3641}{{\em Phys.
  Rev.} {\bfseries D48} (1993) 3641--3647},
\href{http://arxiv.org/abs/gr-qc/9211017}{{\ttfamily arXiv:gr-qc/9211017}}.
%%CITATION = GR-QC/9211017;%%.

\bibitem{Bekenstein:2004ne}
J.~D. Bekenstein, ``{Relativistic gravitation theory for the MOND paradigm},''
  \href{http://dx.doi.org/10.1103/PhysRevD.70.083509}{{\em Phys. Rev.}
  {\bfseries D70} (2004) 083509},
\href{http://arxiv.org/abs/astro-ph/0403694}{{\ttfamily
  arXiv:astro-ph/0403694}}.
%%CITATION = ASTRO-PH/0403694;%%.

\bibitem{Zumalacarregui:2010wj}
M.~Zumalacarregui, T.~S. Koivisto, D.~F. Mota, and P.~Ruiz-Lapuente,
  ``{Disformal Scalar Fields and the Dark Sector of the Universe},''
  \href{http://dx.doi.org/10.1088/1475-7516/2010/05/038}{{\em JCAP} {\bfseries
  1005} (2010) 038},
\href{http://arxiv.org/abs/1004.2684}{{\ttfamily arXiv:1004.2684
  [astro-ph.CO]}}.
%%CITATION = 1004.2684;%%.

\bibitem{Steigman:2010zz}
G.~Steigman, ``{Primordial Nucleosynthesis: The Predicted and Observed
  Abundances and Their Consequences},''
\href{http://arxiv.org/abs/1008.4765}{{\ttfamily arXiv:1008.4765
  [astro-ph.CO]}}.
%%CITATION = 1008.4765;%%.

\bibitem{Izotov:2010ca}
Y.~I. Izotov and T.~X. Thuan, ``{The primordial abundance of 4He: evidence for
  non-standard big bang nucleosynthesis},''
  \href{http://dx.doi.org/10.1088/2041-8205/710/1/L67}{{\em Astrophys. J.}
  {\bfseries 710} (2010) L67--L71},
\href{http://arxiv.org/abs/1001.4440}{{\ttfamily arXiv:1001.4440
  [astro-ph.CO]}}.
%%CITATION = 1001.4440;%%.

\bibitem{GonzalezGarcia:2010un}
M.~C. Gonzalez-Garcia, M.~Maltoni, and J.~Salvado, ``{Robust Cosmological
  Bounds on Neutrinos and their Combination with Oscillation Results},''
  \href{http://dx.doi.org/10.1007/JHEP08(2010)117}{{\em JHEP} {\bfseries 08}
  (2010) 117},
\href{http://arxiv.org/abs/1006.3795}{{\ttfamily arXiv:1006.3795 [hep-ph]}}.
%%CITATION = 1006.3795;%%.

\bibitem{Hamann:2010bk}
J.~Hamann, S.~Hannestad, G.~G. Raffelt, I.~Tamborra, and Y.~Y.~Y. Wong,
  ``{Cosmology seeking friendship with sterile neutrinos},''
  \href{http://dx.doi.org/10.1103/PhysRevLett.105.181301}{{\em Phys. Rev.
  Lett.} {\bfseries 105} (2010) 181301},
\href{http://arxiv.org/abs/1006.5276}{{\ttfamily arXiv:1006.5276 [hep-ph]}}.
%%CITATION = 1006.5276;%%.

\bibitem{Einstein:1945id}
A.~Einstein and E.~G. Straus, ``{The influence of the expansion of space on the
  gravitation fields surrounding the individual stars},''
\href{http://dx.doi.org/10.1103/RevModPhys.17.120}{{\em Rev. Mod. Phys.}
  {\bfseries 17} (1945) 120--124}.
%%CITATION = RMPHA,17,120;%%.

\bibitem{Balbinot:1988zc}
R.~Balbinot, R.~Bergamini, and A.~Comastri, ``{Solution of the Einstein-Strauss
  problem with a Lambda term},''
\href{http://dx.doi.org/10.1103/PhysRevD.38.2415}{{\em Phys. Rev.} {\bfseries
  D38} (1988) 2415--2418}.
%%CITATION = PHRVA,D38,2415;%%.

\bibitem{Barausse:2007pn}
E.~Barausse, T.~P. Sotiriou, and J.~C. Miller, ``{A no-go theorem for
  polytropic spheres in Palatini f(R) gravity},''
  \href{http://dx.doi.org/10.1088/0264-9381/25/6/062001}{{\em Class. Quant.
  Grav.} {\bfseries 25} (2008) 062001},
\href{http://arxiv.org/abs/gr-qc/0703132}{{\ttfamily arXiv:gr-qc/0703132}}.
%%CITATION = GR-QC/0703132;%%.

\end{thebibliography}\endgroup
}
\end{document}